\documentclass[twocolumn]{aastex7}
\usepackage{multirow}
\usepackage{amsmath}
\usepackage{amssymb}
\usepackage{balance}
\usepackage{lmodern}
\newcommand{\nhe}{$\log\,(N_{\rm He}/N_{\rm H})$}
\newcommand{\teff}{$T_{\rm eff}$}
\newcommand{\Msun}{$\rm M_{\odot}$}

\begin{document}

\title{Observational Parameters of Blue Large-amplitude Pulsators\footnote{Based on photometric observations obtained with the 1.3-m Warsaw telescope and spectroscopic data collected with the 6.5-m Magellan-Baade telescope at the Las Campanas Observatory of the Carnegie Institution for Science}}

\author[0000-0002-2339-5899]{Pawe{\l} Pietrukowicz}
\affiliation{Astronomical Observatory, University of Warsaw, Al. Ujazdowskie 4, 00-478 Warszawa, Poland}
\email[show]{pietruk@astrouw.edu.pl}

\author[0000-0002-7547-6180]{Marilyn Latour}
\affiliation{Institute for Astrophysics and Geophysics, Georg-August-University G\"ottingen, Friedrich-Hund-Platz 1, D-37077 G\"ottingen, Germany}
\email[show]{marilyn.latour@uni-goettingen.de}

\author[0000-0002-7777-0842]{Igor Soszy\'nski}
\affiliation{Astronomical Observatory, University of Warsaw, Al. Ujazdowskie 4, 00-478 Warszawa, Poland}
\email{soszynsk@astrouw.edu.pl}

\author[0000-0003-0483-5083]{Francesco Di Mille}
\affiliation{Las Campanas Observatory, Carnegie Institution for Science, Colina el Pino, Casilla 601, La Serena, Chile}
\email{fdimille@carnegiescience.edu}

\author[0000-0000-0000-0000]{Piera Soto King}
\affiliation{Departamento de F{\'i}sica, Universidad de La Serena, Av. Cisternas 1200, La Serena, Chile}
\email{psoto@userena.cl}

\author[0000-0001-7978-7077]{Rodolfo Angeloni}
\affiliation{Gemini Observatory / NSF$'$s NOIRLab, Casilla 603, La Serena, Chile}
\email{rodolfo.angeloni@noirlab.edu}

\author[0000-0002-9245-6368]{Rados{\l}aw Poleski}
\affiliation{Astronomical Observatory, University of Warsaw, Al. Ujazdowskie 4, 00-478 Warszawa, Poland}
\email{rpoleski@astrouw.edu.pl}

\author[0000-0001-5207-5619]{Andrzej Udalski}
\affiliation{Astronomical Observatory, University of Warsaw, Al. Ujazdowskie 4, 00-478 Warszawa, Poland}
\email{udalski@astrouw.edu.pl}

\author[0000-0002-0548-8995]{Micha{\l} K. Szyma\'nski}
\affiliation{Astronomical Observatory, University of Warsaw, Al. Ujazdowskie 4, 00-478 Warszawa, Poland}
\email{msz@astrouw.edu.pl}

\author[0000-0001-6364-408X]{Krzysztof Ulaczyk}
\affiliation{Department of Physics, University of Warwick, Coventry CV4 7AL, UK}
\affiliation{Astronomical Observatory, University of Warsaw, Al. Ujazdowskie 4, 00-478 Warszawa, Poland}
\email{kulaczyk@astrouw.edu.pl}

\author[0000-0003-4084-880X]{Szymon Koz{\l}owski}
\affiliation{Astronomical Observatory, University of Warsaw, Al. Ujazdowskie 4, 00-478 Warszawa, Poland}
\email{simkoz@astrouw.edu.pl}

\author[0000-0002-2335-1730]{Jan Skowron}
\affiliation{Astronomical Observatory, University of Warsaw, Al. Ujazdowskie 4, 00-478 Warszawa, Poland}
\email{jskowron@astrouw.edu.pl}

\author[0000-0001-9439-604X]{Dorota M. Skowron}
\affiliation{Astronomical Observatory, University of Warsaw, Al. Ujazdowskie 4, 00-478 Warszawa, Poland}
\email{dszczyg@astrouw.edu.pl}

\author[0000-0001-7016-1692]{Przemek Mr\'oz}
\affiliation{Astronomical Observatory, University of Warsaw, Al. Ujazdowskie 4, 00-478 Warszawa, Poland}
\email{pmroz@astrouw.edu.pl}

\author[0000-0002-9326-9329]{Krzysztof Rybicki}
\affiliation{Astronomical Observatory, University of Warsaw, Al. Ujazdowskie 4, 00-478 Warszawa, Poland}
\affiliation{Department of Particle Physics and Astrophysics, Weizmann Institute of Science, Rehovot 76100, Israel}
\email{krybicki@astrouw.edu.pl}

\author[0000-0002-6212-7221]{Patryk Iwanek}
\affiliation{Astronomical Observatory, University of Warsaw, Al. Ujazdowskie 4, 00-478 Warszawa, Poland}
\email{piwanek@astrouw.edu.pl}

\author[0000-0002-3051-274X]{Marcin Wrona}
\affiliation{Astronomical Observatory, University of Warsaw, Al. Ujazdowskie 4, 00-478 Warszawa, Poland}
\affiliation{Department of Astrophysics and Planetary Science, Villanova University, 800 East Lancaster Avenue, Villanova, PA 19085, USA}
\email{mwrona@astrouw.edu.pl}

\author[0000-0002-1650-1518]{Mariusz Gromadzki}
\affiliation{Astronomical Observatory, University of Warsaw, Al. Ujazdowskie 4, 00-478 Warszawa, Poland}
\email{marg@astrouw.edu.pl}



\begin{abstract}

Blue large-amplitude pulsators (BLAPs) are a recently discovered class of short-period pulsating variable stars. In this work, we present new information on these stars based on photometric and spectroscopic data obtained for known and new objects detected by the Optical Gravitational Lensing Experiment (OGLE) survey. BLAPs are evolved objects with pulsation periods in the range of 3--75~minutes, stretching between subdwarf B-type stars and upper main-sequence stars in the Hertzsprung-Russell diagram. In general, BLAPs are single-mode stars pulsating in the fundamental radial mode. Their phase-folded light curves are typically sawtooth-shaped, but many longer-period objects exhibit an additional bump. The long-term OGLE observations show that the period change rates of BLAPs are usually of the order of $10^{-7}$~yr$^{-1}$ and in a quarter of the sample are negative. The spectroscopic data indicate that the BLAPs form a homogeneous group in the period, surface gravity, and effective temperature spaces. However, we observe a split into two groups in terms of helium-to-hydrogen content. The atmospheres of the He-enriched BLAPs are more abundant in metals (about 5~times) than the atmosphere of the Sun. We discover that BLAPs obey a period--gravity relationship and we use the distance to OGLE-BLAP-009 to derive a period--luminosity relation. Most of the stars observed in the OGLE Galactic bulge fields seem to reside in the bulge, while the remaining objects likely are in the foreground Galactic disk.

\end{abstract}


\keywords{\uat{Blue large-amplitude pulsators}{2112} --- \uat{Pulsating variable stars}{1307}}


\section{Introduction} \label{sec:intro}

Blue large-amplitude Pulsators (BLAPs) form a recently discovered class of objects in the realm of variable stars. The first object was serendipitously found during a variability search of a 7~deg$^2$ area containing nine million stars of the Galactic disk in the direction tangent to the Carina arm \citep{2013AcA....63..379P}. The area was monitored during the third phase of the Optical Gravitational Lensing Experiment \citep[OGLE-III, years 2001--2009,][]{2008AcA....58...69U} in a quest for planetary transits \citep[e.g.,][]{2002AcA....52..317U,2004AcA....54..313U}. With an amplitude of almost 0.28 mag in the $I$ band, a period of 28.25 minutes, and a sawtooth-shaped light curve, the star was initially classified as a pulsating variable of $\delta$~Sct type. However, such a high amplitude at such a short period was very unusual for this type of pulsator. A low-resolution spectrum \citep{2015AcA....65...63P} showed that the mysterious object is of late O type, thus it is much hotter than $\delta$~Sct stars covering spectral types from A0 to F6. The estimated surface gravity was stronger than in upper main-sequence stars and weaker than that in hot subdwarfs, which ruled out the possibility that it is a $\beta$~Cep variable or an oscillating hot subdwarf.

The detection of 13 variable stars with very similar photometric properties in the OGLE Galactic bulge fields and the results from the analysis of spectra obtained for several objects allowed \citet{2017NatAs...1E.166P} to introduce the new class of variable stars. The following characteristics of the stars gave their name to the class, blue large-amplitude pulsators: (1) location blueward of the main sequence in the color-magnitude diagram, (2) exceptionally large amplitudes in optical passbands for such hot objects, and (3) proof of a pulsation nature for the observed brightness variations (color and effective temperature changes over the cycle). The published variables have amplitudes between 0.19 and 0.36 mag in the Cousins $I$ band (0.22--0.43 mag in the Johnson $V$ band) and pulsation periods between 22 and 40 minutes. All objects but the prototype, OGLE-BLAP-001, seemed to be single-mode pulsators. Long-term OGLE data showed that the detected variables are stable over time with period change rates of the order of several times $10^{-7}$ yr$^{-1}$. In most cases, the changes are positive. Information obtained from spectroscopy supported the view of BLAPs as a very uniform group of stars, with surface gravity $\log g$ in the range of 4.2--4.6 (where $g$ is in units of cm~s$^{-2}$), effective temperature \teff\ from about 26,000 to 32,000 K, and a helium-to-hydrogen ratio \nhe\ from $-0.55$ to $-0.40$ dex.

In the following years, new BLAPs were discovered. Four variables with much shorter periods than in the original pulsators, between 3 and 8 minutes, were found in high-cadence data from the Zwicky Transient Facility \citep[ZTF;][]{2019PASP..131a8002B} by \citet{2019ApJ...878L..35K}. The phase-folded light curves of these variables are more symmetric than those of the original BLAPs. Their amplitudes reach 0.30 mag in the ZTF-$r$ filter. Dedicated spectroscopic observations showed that the surface gravity of the stars is about an order of magnitude stronger ($\log g$ between 5.2 and 5.7 dex). Therefore, the objects were named high-gravity BLAPs. They are on average slightly hotter (\teff\ from 31,400 to 34,000 K) than the original pulsators and their helium abundance turned out to be much lower (\nhe\ between $-2.4$ and $-2.0$ dex).

Recently, a detection of 22 candidates for BLAPs, including six high-gravity pulsators, has been reported by \citet{2022MNRAS.511.4971M}. The stars were found by combining photometric measurements from the Gaia space mission \citep{2018A&A...616A...1G} with time-series data from ZTF. Their periods range from about 2.4 to 55 minutes. Seven of the candidates are located at Galactic latitudes higher than $10\degr$, while for three of them $|b|>30\degr$. One of the objects, ZGP-BLAP-01 with a pulsation period of 18.93 minutes, also known as TMTS-BLAP-1, has been found to have an unusually large period change rate of $2.2\times10^{-6}$ yr$^{-1}$ \citep{2023NatAs...7..223L}. Based on data from the OmegaWhite survey \citep{2015MNRAS.454..507M}, \citet{2022MNRAS.513.2215R} identified four BLAPs toward the bulge, one of which was already known. The periods of these stars are between 10 and 33 minutes. Very recently, a discovery of a new low-gravity BLAP in the data from the SkyMapper Southern Sky Survey was announced by \cite{2024MNRAS.529.1414C}.

Finally, the first two BLAPs in binary systems have been found. \citet{2022A&A...663A..62P} discovered a 32.37 minute pulsator that orbits a main-sequence B-type star every 23.08 days. Interestingly, the system (HD 133729) is located at a moderately high Galactic latitude of almost $+23\degr$. The previously mentioned object TMTS-BLAP-1 probably forms a wide binary of an orbital period of 1576 days \citep{2023NatAs...7..223L}. The number of known BLAPs has increased by 20 new objects discovered in the OGLE Galactic disk fields \citep{2023AcA....73....1B}. At the moment of writing (2024 January), the total number of BLAPs exceeds 80 (if we include the stars reported in this work).

Since the discovery of the first BLAPs, several theoretical attempts have been undertaken to explain the nature and evolutionary status of these stars. It was already known that such large-amplitude oscillations may occur in stars with inflated envelopes or giant-like structure \citep{1977AcA....27...95D}. BLAPs are thought to pulsate through the $\kappa$-mechanism due to the presence of an enhanced amount of iron-group elements in the outer layers (the so called Z-bump in the opacity). Sufficient accumulation of iron and nickel around the opacity maximum at a temperature of about $2\times10^5$~K is possible thanks to atomic diffusion and radiative levitation \citep{2018MNRAS.481.3810B,2020MNRAS.492..232B}.

Three stellar structure models have been proposed in the literature. In the first model, BLAPs are low-mass ($M\approx0.3$ \Msun) helium-core, shell hydrogen-burning pre-white dwarfs (pre-WDs) approaching the white dwarf cooling track \citep{2018arXiv180907451C,2018MNRAS.477L..30R}. Another possibility is that the stars are core helium-burning postgiants evolving toward the extended horizontal branch \citep{2018MNRAS.481.3810B,2018MNRAS.478.3871W}. In the third model, BLAPs are in a phase of stable shell helium-burning on the surface of the CO core when the star evolves off the extended horizontal branch \citep{2019ApJ...878L..35K,2023NatAs...7..223L}. The latter two models imply more massive stars ($M>0.45$ \Msun).

The formation of BLAPs is still a mystery. To create such a structure, the star needs to be stripped of its hydrogen-rich envelope after the helium core has formed. It is possible through mass transfer and/or common envelope ejection in binary system evolution \citep{2021MNRAS.507..621B}. \citet{2020ApJ...903..100M} proposed that BLAPs could be the surviving companions of single-degenerate Type Ia supernovae. In their original paper, \citet{2017NatAs...1E.166P} suggested that a significant mass loss could take place during the encounter of the star with the Milky Way's central supermassive black hole. A lack of observed BLAPs in the Magellanic Clouds \citep{2018pas6.conf..258P} would support this hypothesis.

In this paper, we present new data on BLAPs that will help to understand the nature and origin of these stars. We report the detection of 23 additional stars in the OGLE-IV inner Galactic bulge area, six of which are in the period range of 14--20 minutes. For some of the stars, we obtained moderate-resolution spectra. We show that the high-gravity BLAPs and the original BLAPs form a single class of pulsators. We announce the discovery of a period--gravity relationship for BLAPs, the first such relation for Z-bump pulsators. 

Our paper is organized as follows. We describe the photometric and spectroscopic data in Section~\ref{sec:data}. Section~\ref{sec:analysis} is devoted to the analysis of the data. In the broad Section~\ref{sec:results}, we present the photometric properties of BLAPs, including information on multimodality and period changes. We also derive the atmospheric parameters of the metal composition of a handful of BLAPs. Further, we announce the discovery of the period-gravity relation and we derive the first period-luminosity ($P$-$L$) relation for BLAPs. At the end of this section, we focus on the kinematics and location of these stars in the Milky Way. Finally, we discuss and summarize our results in Section~\ref{sec:disc} and Section~\ref{sec:conc}, respectively.

\section{Data} \label{sec:data}

\begin{figure*}
\centering
\includegraphics[width=0.98\textwidth]{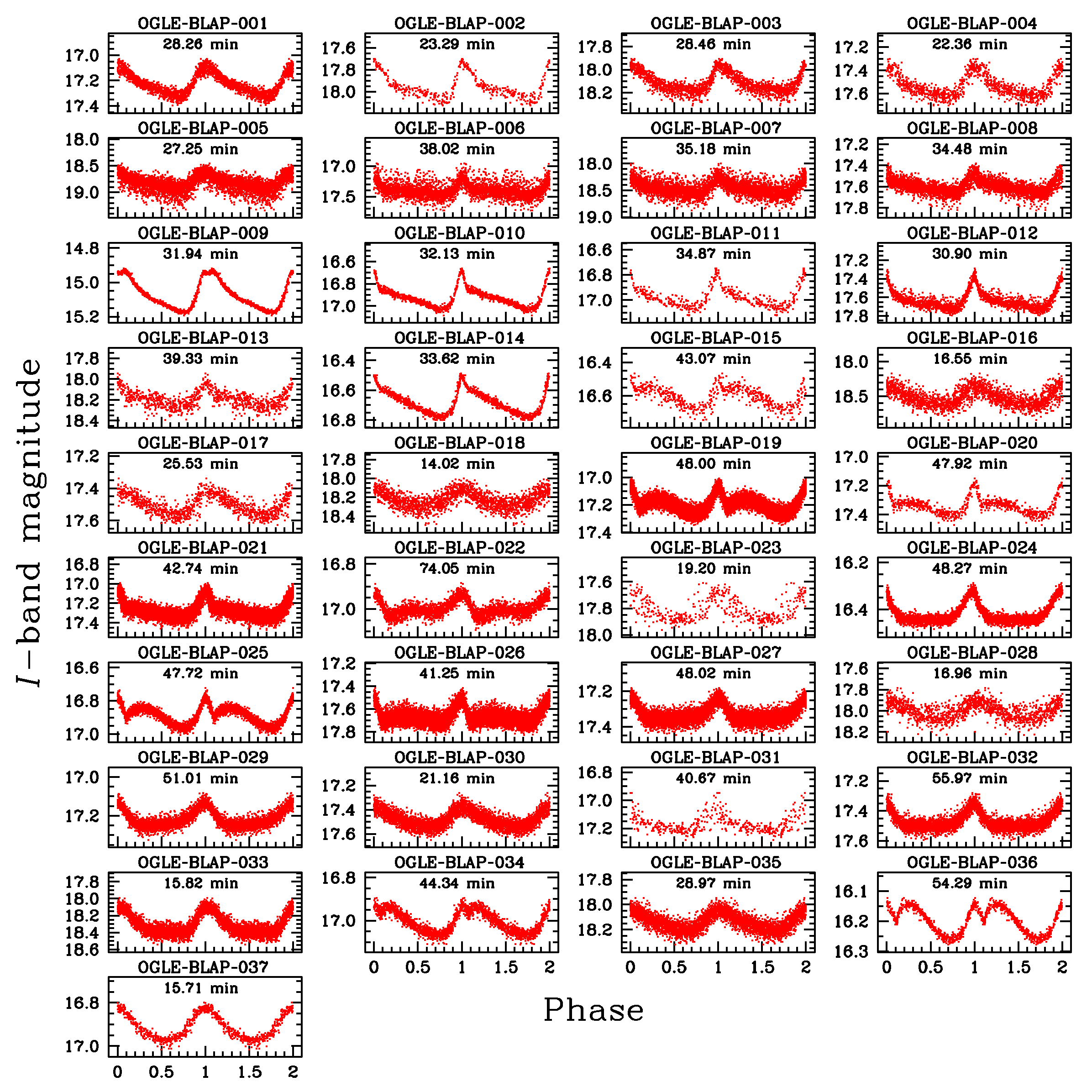}
\caption{Phase-folded $I$-band light curves of the prototype object OGLE-BLAP-001 and 36 BLAPs detected by the OGLE survey toward the Galactic bulge. For majority of the bulge objects, all available OGLE-IV time-series data (from years 2010--2019) are presented, whereas for several objects only part of the data was selected to clearly show the light curve shape (e.g., OGLE-BLAP-030). The most characteristic feature in BLAPs is the sharp maximum. An additional bump is seen in objects with periods exceeding 40 minutes.}
\label{fig:curves}
\end{figure*}

\subsection{Photometric Data}

The photometric data presented in this paper were obtained by OGLE. OGLE is a long-term survey conducted with the 1.3 m Warsaw telescope at the Las Campanas Observatory, Chile\footnote{The observatory is operated by the Carnegie Institution for Science.}. The survey started in 1992, with the primary goal to detect microlensing events through monitoring of dense stellar fields. Over years 1992--1995, OGLE observed a 0.9 deg$^2$ region around Baade's Window toward the Galactic bulge. A larger bulge area covering 11 deg$^2$ was monitored during the second phase in years 1997--2000 \citep[OGLE-II,][]{1997AcA....47..319U}. In that phase, regular observations of the Magellanic Clouds began. The third phase started with the installation of an eight-detector mosaic camera, which allowed the project to increase the surveyed area again. During OGLE-III \citep{2008AcA....58...69U}, conducted between 2001 June and 2009 May, the bulge coverage reached about 69 deg$^2$. The fourth phase \citep[OGLE-IV,][]{2015AcA....65....1U} was launched in 2010 March and continues to the present except for a break due to the Covid-19 pandemic, when regular observations were stopped for 29 months, between 2020 March and 2022 August. The OGLE-IV camera is a 32-detector mosaic covering a field of view of 1.4 deg$^2$. Currently, the survey monitors up to 3,600 deg$^2$ of the sky including the Magellanic System and the entire Milky Way stripe visible from the Las Campanas Observatory. Images are collected mainly through the $I$ filter. To secure color information, additional images are taken through the $V$ filter. The data is reduced with the difference image analysis \citep{2000AcA....50..421W}, a method developed especially for dense stellar fields. Regular, precise, long-term brightness measurements by the OGLE survey have led to the discovery and unambiguous classification of over one million variable objects including transient, irregular, and periodic sources of various types \citep[e.g.,][]{2015AcA....65..313M,2016AcA....66..405S,2019ApJ...879..114I,2019AcA....69..321S,2022ApJS..259...16W}.

\subsection{Spectroscopic Data}

The spectroscopic follow-up data were obtained with MagE, a moderate-resolution echelle spectrograph attached to the 6.5-m Magellan-Baade telescope located at the Las Campanas Observatory. The spectrograph covers a wavelength range from about 3200~{\AA} to 10000~{\AA}. Observations were collected over three nights: on 2018 May 19/20 and 2019 May 31/June 1 under Chilean telescope time (programs CN2018A-102 and CN2019A-13, respectively) for 11 objects toward the Galactic bulge presented in this work, and on 2023 July 30/31 during telescope engineering time for three recently discovered objects in the Galactic disk \citep{2023AcA....73....1B}. During the observations the seeing varied between 0\farcs5 and 0\farcs7. We used a slit width of 1\farcs0 which provided a resolving power of 4100 or spectral resolution of about 1.0 {\AA} at 5000 {\AA}. For all but one object, two consecutive single spectra were taken, each exposed for half of the pulsation cycle. Experimentally, in the case of OGLE-BLAP-033, we took four exposures, each covering a quarter of the cycle. The science exposures were followed by an exposure of the calibration lamp. All the spectra were reduced using the \texttt{IRAF} package\footnote{IRAF was distributed by the National Optical Astronomy Observatory, USA, which is operated by the Association of Universities for Research in Astronomy, Inc., under a cooperative agreement with the National Science Foundation.} \citep{1986SPIE..627..733T,1993ASPC...52..173T}. Debiasing, flat-fielding, and wavelength calibrations were performed in the standard way. For the wavelength calibrations, a Th-Ar lamp was used. In this paper, we also refer to the results of spectroscopic observations of the prototype object from the Galactic disk, OGLE-BLAP-001, taken with MagE on the night of 2016 April 22 using the same strategy (two science spectra followed by the calibration lamp). Thus, our spectroscopic sample includes 15 OGLE BLAPs observed with the MagE spectrograph. Eleven of them are from the Galactic bulge fields (this work), while the remaining four stars are from the Galactic disk fields \citep{2017NatAs...1E.166P,2023AcA....73....1B}. All the spectra are presented in normalized flux.

\begin{sidewaystable*}[h!]
\centering \caption{Photometric Data on the Prototype BLAP and BLAPs Detected toward the Galactic Bulge}
{\fontsize{7}{8} \selectfont
\begin{tabular}{cccccccccccc}
\hline
Name          & R.A.(2000.0)  & Decl.(2000.0)   &    $l$    &    $b$     & $<I>$  & $<V>$  & $A_I$ & $A_V$ &  $P$   &         $r$       & Remarks \\
              & (hh:mm:ss)     & (deg:arcmin:arcsec) & ($\degr$) & ($\degr$)  & (mag)  & (mag)  & (mag) & (mag) & (minutes)  & ($10^{-7}$ yr$^{-1}$) & \\
\hline
OGLE-BLAP-001 & 10:41:48.77 & $-61$:25:08.5 & 288.06355 & $-2.34712$ & 17.223 & 17.706 & 0.236 & 0.411 & 28.255 & $ -1.65 \pm 0.31$ & triplet \\
OGLE-BLAP-002 & 17:43:58.02 & $-19$:16:54.1 &   8.05922 & $ 5.34683$ & 17.951 & 18.893 & 0.357 & 0.33  & 23.286 & $-$               & \\
OGLE-BLAP-003 & 17:44:51.48 & $-24$:10:04.0 &   3.98204 & $ 2.63080$ & 18.111 & 19.170 & 0.235 & 0.266 & 28.458 & $ +0.81 \pm 0.22$ & \\
OGLE-BLAP-004 & 17:51:04.72 & $-22$:09:03.4 &   6.44968 & $ 2.44527$ & 17.534 & 18.861 & 0.259 & 0.429 & 22.357 & $-$               & \\
OGLE-BLAP-005 & 17:52:18.73 & $-31$:56:35.0 & 358.16325 & $-2.78388$ & 18.818 & 20.000 & 0.302 & 0.332 & 27.253 & $ +0.78 \pm 0.16$ & \\
OGLE-BLAP-006 & 17:55:02.88 & $-29$:50:37.5 &   0.27313 & $-2.22853$ & 17.398 & 18.349 & 0.308 & 0.358 & 38.015 & $ -3.32 \pm 0.37$ & \\
OGLE-BLAP-007 & 17:55:57.52 & $-28$:52:11.0 &   1.21533 & $-1.90988$ & 18.455 & 19.276 & 0.284 & 0.296 & 35.182 & $ -3.41 \pm 0.45$ & \\
OGLE-BLAP-008 & 17:56:48.26 & $-32$:21:35.6 & 358.28172 & $-3.81528$ & 17.598 & 18.732 & 0.174 & 0.271 & 34.481 & $ +2.79 \pm 0.18$ & \\
OGLE-BLAP-009 & 17:58:48.20 & $-27$:16:53.7 &   2.90511 & $-1.65832$ & 15.070 & 15.639 & 0.242 & 0.275 & 31.935 & $ +1.50 \pm 0.05$ & dip \\
OGLE-BLAP-010 & 17:58:59.22 & $-35$:18:07.0 & 355.94188 & $-5.66572$ & 16.924 & 17.260 & 0.343 & 0.383 & 32.133 & $ +0.47 \pm 0.19$ & \\
OGLE-BLAP-011 & 18:00:23.24 & $-35$:58:03.1 & 355.49826 & $-6.23954$ & 16.977 & 17.260 & 0.286 & 0.23  & 34.875 & $-$               & \\
OGLE-BLAP-012 & 18:05:44.20 & $-30$:11:15.2 &   1.11870 & $-4.41006$ & 17.625 & 18.251 & 0.341 & 0.353 & 30.897 & $ +0.05 \pm 0.10$ & \\
OGLE-BLAP-013 & 18:05:52.70 & $-26$:48:18.0 &   4.09671 & $-2.79459$ & 18.194 & 19.251 & 0.228 & 0.236 & 39.326 & $ +8.20 \pm 0.50$ & \\
OGLE-BLAP-014 & 18:12:41.79 & $-31$:12:07.8 &   0.94058 & $-6.20906$ & 16.682 & 16.788 & 0.265 & 0.342 & 33.623 & $ +4.83 \pm 0.39$ & \\
OGLE-BLAP-015 & 17:39:58.35 & $-22$:37:22.5 &   4.71127 & $ 4.39525$ & 16.602 & 17.525 & 0.159 & 0.206 & 43.073 & $ +1.91 \pm 0.99$ & bump \\
OGLE-BLAP-016 & 17:47:41.33 & $-35$:14:27.1 & 354.83143 & $-3.64179$ & 18.511 & 19.024 & 0.260 & 0.265 & 16.548 & $ +0.41 \pm 0.19$ & \\
OGLE-BLAP-017 & 17:48:01.46 & $-21$:20:43.2 &   6.77880 & $ 3.46702$ & 17.508 & 18.720 & 0.149 & 0.192 & 25.535 & $-$               & \\
OGLE-BLAP-018 & 17:49:44.57 & $-34$:54:31.7 & 355.33301 & $-3.83247$ & 18.223 & 18.510 & 0.179 & 0.193 & 14.022 & $ +0.40 \pm 0.15$ & \\
OGLE-BLAP-019 & 17:53:34.39 & $-30$:02:11.6 & 359.94513 & $-2.05015$ & 17.196 & 18.386 & 0.191 & 0.227 & 48.005 & $ +2.05 \pm 0.14$ & bump \\
OGLE-BLAP-020 & 17:56:32.05 & $-20$:37:15.7 &   8.41333 & $ 2.12352$ & 17.341 & 18.618 & 0.225 & 0.348 & 47.923 & $-$               & bump \\
OGLE-BLAP-021 & 17:58:30.23 & $-29$:38:09.2 &   0.82920 & $-2.77380$ & 17.265 & 17.950 & 0.261 & 0.306 & 42.742 & $ -5.24 \pm 0.26$ & \\
OGLE-BLAP-022 & 17:58:51.29 & $-28$:28:53.3 &   1.86967 & $-2.26578$ & 17.007 & 18.041 & 0.087 & 0.107 & 74.052 & $ +6.16 \pm 0.54$ & bump \\
OGLE-BLAP-023 & 17:58:57.27 & $-36$:33:41.5 & 354.83492 & $-6.27767$ & 17.807 & 17.952 & 0.20  & 0.20  & 19.196 & $-$               & \\
OGLE-BLAP-024 & 17:59:43.70 & $-28$:44:26.4 &   1.73988 & $-2.56094$ & 16.418 & 17.167 & 0.124 & 0.186 & 48.268 & $ -3.53 \pm 0.26$ & \\
OGLE-BLAP-025 & 17:59:47.33 & $-33$:07:07.6 & 357.93256 & $-4.73645$ & 16.886 & 17.744 & 0.182 & 0.227 & 47.724 & $ +2.49 \pm 0.74$ & bump \\
OGLE-BLAP-026 & 18:00:12.90 & $-27$:34:22.2 &   2.80805 & $-2.07490$ & 17.662 & 18.664 & 0.209 & 0.246 & 41.250 & $ +0.27 \pm 0.19$ & bump \\
OGLE-BLAP-027 & 18:00:30.66 & $-27$:36:25.1 &   2.81090 & $-2.14879$ & 17.328 & 19.011 & 0.119 & 0.167 & 48.018 & $ +0.46 \pm 0.22$ & \\
OGLE-BLAP-028 & 18:02:38.90 & $-31$:25:44.2 & 359.70732 & $-4.43447$ & 17.987 & 18.644 & 0.177 & 0.204 & 16.965 & $ -5.52 \pm 0.32$ & \\
OGLE-BLAP-029 & 18:04:12.05 & $-28$:02:54.9 &   2.82788 & $-3.07627$ & 17.214 & 17.931 & 0.113 & 0.124 & 51.008 & $ -4.42 \pm 0.43$ & \\
OGLE-BLAP-030 & 18:04:57.10 & $-28$:08:06.2 &   2.83324 & $-3.26300$ & 17.477 & 17.954 & 0.174 & 0.220 & 21.162 & $+45.80 \pm 0.65$ & triplemode \\
OGLE-BLAP-031 & 18:05:56.03 & $-35$:53:55.5 & 356.10354 & $-7.19029$ & 17.170 & 17.203 & 0.31  & 0.18  & 40.671 & $-$               & \\
OGLE-BLAP-032 & 18:06:28.93 & $-26$:31:10.4 &   4.41210 & $-2.77361$ & 17.467 & 18.560 & 0.152 & 0.168 & 55.968 & $ +1.98 \pm 0.26$ & \\
OGLE-BLAP-033 & 18:07:31.70 & $-27$:58:20.8 &   3.25208 & $-3.68113$ & 18.300 & 18.907 & 0.301 & 0.349 & 15.822 & $ +0.28 \pm 0.06$ & \\
OGLE-BLAP-034 & 18:07:50.55 & $-30$:04:47.6 &   1.43366 & $-4.75695$ & 17.005 & 17.696 & 0.130 & 0.185 & 44.335 & $ +2.27 \pm 0.62$ & bump \\
OGLE-BLAP-035 & 18:10:38.58 & $-25$:16:08.6 &   5.96101 & $-2.98878$ & 18.135 & 19.122 & 0.157 & 0.191 & 28.975 & $ +1.20 \pm 0.68$ & \\
OGLE-BLAP-036 & 18:10:45.47 & $-32$:28:21.0 & 359.61805 & $-6.44446$ & 16.199 & 16.539 & 0.120 & 0.145 & 54.289 & $ +2.85 \pm 0.66$ & bump \\
OGLE-BLAP-037 & 18:19:20.89 & $-27$:29:56.3 &   4.91343 & $-5.76037$ & 16.909 & 17.207 & 0.149 & 0.174 & 15.712 & $-$               & \\
\hline
\label{tab:photo}
\end{tabular}}
\end{sidewaystable*}

\section{Analysis}\label{sec:analysis}

\subsection{Photometric Data}

The BLAPs reported in this work were found as a result of a search for short-period ($P<1$~hr) variable objects in the OGLE-IV fields for the inner Galactic bulge. The search was conducted using time-series $I$-band data covering seasons\footnote{Each bulge season starts in February and ends in October.} 2010--2015 with the help of the FNPEAKS code\footnote{\scriptsize http://helas.astro.uni.wroc.pl/deliverables.php?lang=en\&active=fnpeaks}. It calculates Fourier amplitude spectra of unequally spaced data composed of a large number of points. The code substantially reduces the computation time for a discreet Fourier transform by coadding correctly phased, low-resolution Fourier transforms of pieces of a large data set interpolated to high resolution. About 400 million stellar sources were inspected. The number of $I$-band measurements taken in the years 2010--2015 was between about 200 and 13,000 per object, depending on the field. In the investigated period range ($P<1$~hr), any daily or seasonal aliases could not hamper the detection of real pulsators. We analyze time-series data from all OGLE-IV seasons before the pandemic (years 2010--2019) and also data from previous OGLE phases if available.

To study the photometric behavior of short-period variables like BLAPs, we converted moments of the brightness measurements from the Heliocentric Julian Date (HJD) to the Barycentric Julian Date (BJD$_{\rm TDB}$). Accurate pulsation periods and their errors were determined with the help of the \texttt{TATRY} code \citep{1996ApJ...460L.107S}. The code employs periodic orthogonal polynomials to fit the data and analysis of variance statistics to evaluate the quality of the fit. We removed evident outlying points in the phased light curves and corrected the periods. In the detected BLAPs, we looked for possible multimodality and calculated period derivatives. Additional periodic signals were searched for after prewhitening of the phased light curves.

The analysis of the first BLAPs showed that the stars are characterized by slow period changes. Here, we investigate the changes based on a larger sample of objects and a longer time baseline. That includes more recently collected measurements from OGLE-IV and also data from previous OGLE phases. For fast-period-changing objects, we determined pulsation periods for each OGLE-IV season separately to follow the changes on shorter timescales.

We note that the uncertainties of single brightness measurements are of about 0.003 mag at $I=14$ mag, about 0.007 mag at $I=16$ mag, and about 0.03 mag at $I=18$ mag, and they reach 0.15 mag for stars with $I=20$ mag. We determine peak-to-peak amplitudes by finding the best Fourier fit to the phased light curve. Noise limits for periodic signals and amplitude uncertainties are of the order mentioned above.

\subsection{Spectroscopic Data}

The atmospheric parameters for the 14 new BLAPs observed with the MagE spectrograph were derived in the same way as for the prototype star OGLE-BLAP-001 \citep{2017NatAs...1E.166P}. A grid of line-blanketed, non-LTE model atmospheres and synthetic spectra computed with the \texttt{TLUSTY} and \texttt{SYNSPEC} codes \citep{2011ascl.soft09021H,2011ascl.soft09022H} was used to simultaneously fit the hydrogen and helium lines visible in the observed spectra \citep{2010AIPC.1273..259B}. This way, we derived the effective temperature (\teff), surface gravity ($\log g$), and helium abundance (\nhe). For each star, we analyzed the combined spectra, meaning that for each star the individual exposures were coadded. Thus we fit the ``average" spectrum over one pulsation cycle. The uncertainties obtained for the atmospheric parameters are the formal errors returned by the $\chi^2$ fitting procedure and are only representative of the quality of the fit. As shown in \citet{2017NatAs...1E.166P} for the prototype, the atmospheric parameters vary during the pulsation cycle. The changes in $T_{\rm eff}$ can be $\pm$~5,000~K and of $\pm0.2$~dex in $\log g$ \citep{2024MNRAS.52710239B}. The uncertainties quoted on the atmospheric parameters must be considered as minimum values. The radial velocity (RV, $v_{\rm r}$) was also derived for each spectral line as part of the fitting procedure. We then computed the average RV (and standard deviation) obtained from a subset of lines that were well reproduced by the best fit. As for the atmospheric parameters, the RV of the spectral lines varies during the pulsation cycle, as a consequence of the expansion and contraction of the stellar atmosphere. The bluer lines are less affected by the RV variations which is why we used only the RV of the lines blueward of He~\textsc{i} $\lambda$5015 to compute the average RV of each star. Depending on the star, the average RV was computed from three to eight lines.

The obtained photometric as well as spectroscopic data on BLAPs are available to the astronomical community through the OGLE webpage at https://ftp.astrouw.edu.pl/ogle/ogle4/OCVS/BLAP/.

\section{Results} \label{sec:results}

\subsection{Photometric Properties} \label{sec:photo}

The first 14 discovered BLAPs were characterized by amplitudes in the range of 0.2--0.4 mag, pulsation periods between 20 and 40 minutes, and RRab-like light curve shapes \citep{2017NatAs...1E.166P}. Our search for short-period variables in the OGLE-IV fields, covering 172 deg$^2$ of the inner Galactic bulge, led to the detection of 23 objects having more or less similar photometric properties. The spectroscopic follow-up observations clearly show that all the objects belong to the same group of stars, i.e., BLAPs, but with a broader range of parameters and a larger variety of light curve shapes (see Section~\ref{sec:atmo}). In Figure~\ref{fig:curves}, we present the phase-folded $I$-band light curves of the prototype object OGLE-BLAP-001 and all BLAPs from the Galactic bulge fields. Six new pulsators have pulsation periods below 20 minutes. They have nearly symmetric light curves with less sharp maxima than objects with $P>20$ minutes. Roughly half of the BLAPs with a period longer than 40 minutes exhibit an additional bump in the light curve, which is also observed in two known pulsating extreme helium stars, BX Cir ($P=2.56$~hr) and V652 Her \citep[$P=2.59$~hr;][]{1999MNRAS.310.1119K}. The bump has a sinusoidal shape with the maximum usually fainter than the main sharp peak. In the whole sample, object OGLE-BLAP-022 has the longest period (74.05 minutes) and the smallest amplitude (0.087 mag in $I$). This object also shows a reverse light curve in comparison to the other stars with a bump. In the light curve of object OGLE-BLAP-031, there is an unusually large scatter around the maximum. Nevertheless, our spectroscopic data indicate that both stars are true BLAPs (see Section~\ref{sec:atmo}). From the light curves presented in Figure~\ref{fig:curves}, we removed evident outlying data points. In the case of several stars in frequently monitored OGLE fields, for transparency, we show only measurements collected during one selected season (from 2013). This includes object OGLE-BLAP-030 which experiences fast period and amplitude changes (investigated in Section~\ref{sec:changes}). We note that changes are also seen in the light curve of OGLE-BLAP-023. In Figure~\ref{fig:curves}, we show all available data for this object from years 2010--2014. 

We note that additional periodic signals were detected only in two objects: OGLE-BLAP-001 and OGLE-BLAP-030. Hence, a significant majority of BLAPs seem to be single-mode pulsators. However, we cannot rule out the possibility that there are additional modes of very low amplitude ($<0.01$ mag), practically undetectable from the ground. The presented light curves of the two multimode BLAPs are phased with their dominant periods.

In Table~\ref{tab:photo}, we provide information on the 37 BLAPs obtained from the OGLE photometry. The 23 new objects are ordered with increasing R.A. and added to the 14 original variables of this type. We continue naming the variables in the format OGLE-BLAP-NNN, where NNN is a consecutive number. Besides the equatorial and galactic coordinates, mean magnitudes and amplitudes in the $I$ and $V$ bands, and pulsation period, we also provide the rate of period change ($r$) and information on multimodality and the presence of a bump or dip in the light curve (see Section~\ref{sec:changes}).

BLAPs are relatively faint objects. In our set of 37 pulsators, none of the stars is brighter than $V=15$ mag and only three stars have $V<17$ mag. The measured amplitudes of BLAPs are in the range 0.08--0.36 mag in the $I$ band and 0.10--0.43 mag in the $V$ band. The median value of the $V$-band to $I$-band amplitude ratio is about 1.2, which is consistent with the pulsation nature of the stars.

\begin{figure}
\centering
\includegraphics[width=0.45\textwidth]{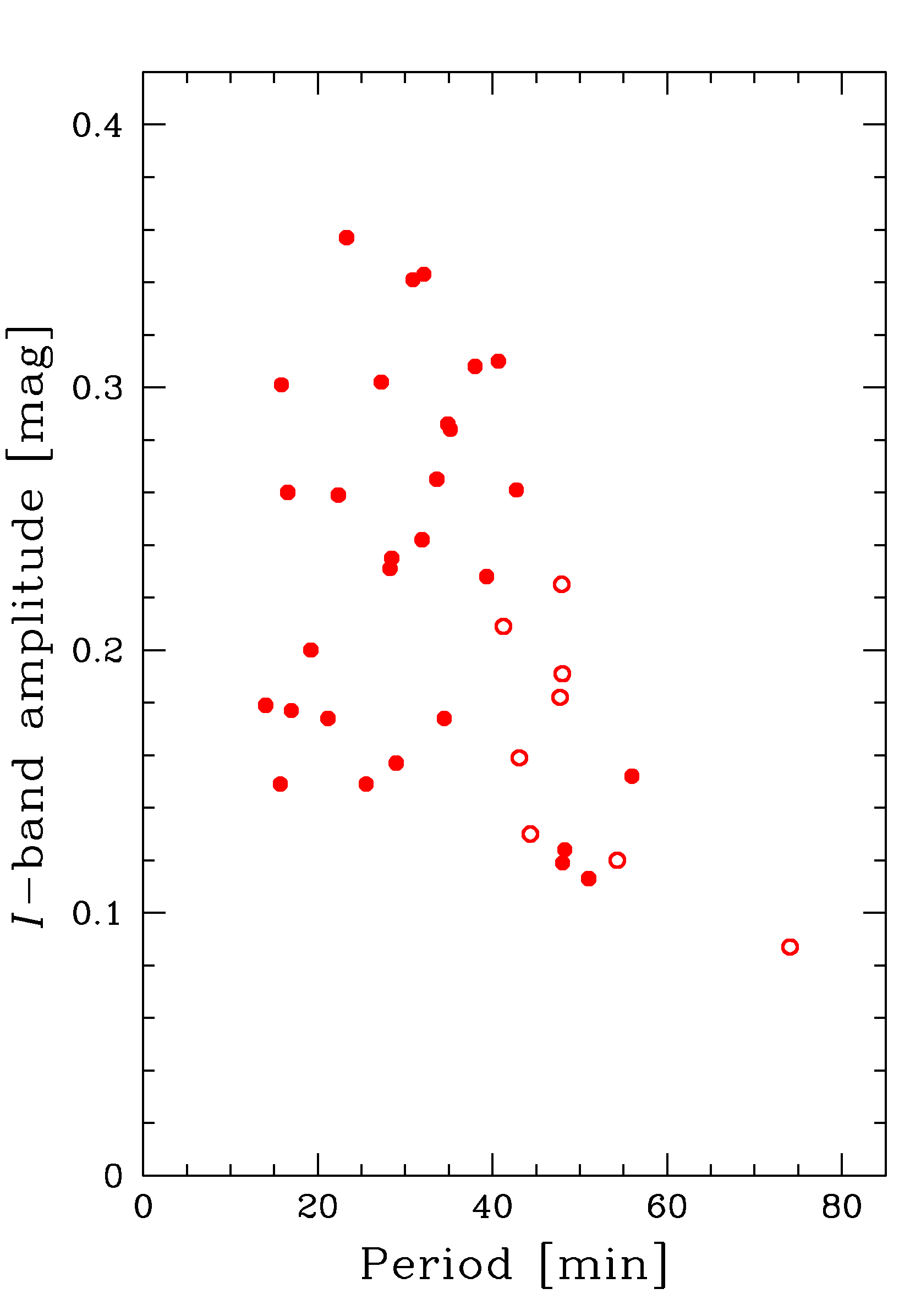}
\caption{Period vs. $I$-band amplitude diagram for BLAPs observed toward the Galactic bulge. The diagram also includes the prototype object from the Galactic disk. Open dots refer to objects with an additional bump in the light curve.}
\label{fig:peramp}
\end{figure}

The pulsation periods of the investigated BLAPs are between 14 and 74 minutes. In Figure~\ref{fig:peramp}, we show a period-amplitude diagram for the $I$-band data. BLAPs with periods longer than 50 minutes have amplitudes below 0.2 mag. We mark the position of the eight stars with a bump in their light curve. Two of these stars have amplitudes above 0.2 mag.

\subsection{Multimodality and Period Changes} \label{sec:changes}

\begin{figure}
\centering
\includegraphics[width=0.45\textwidth]{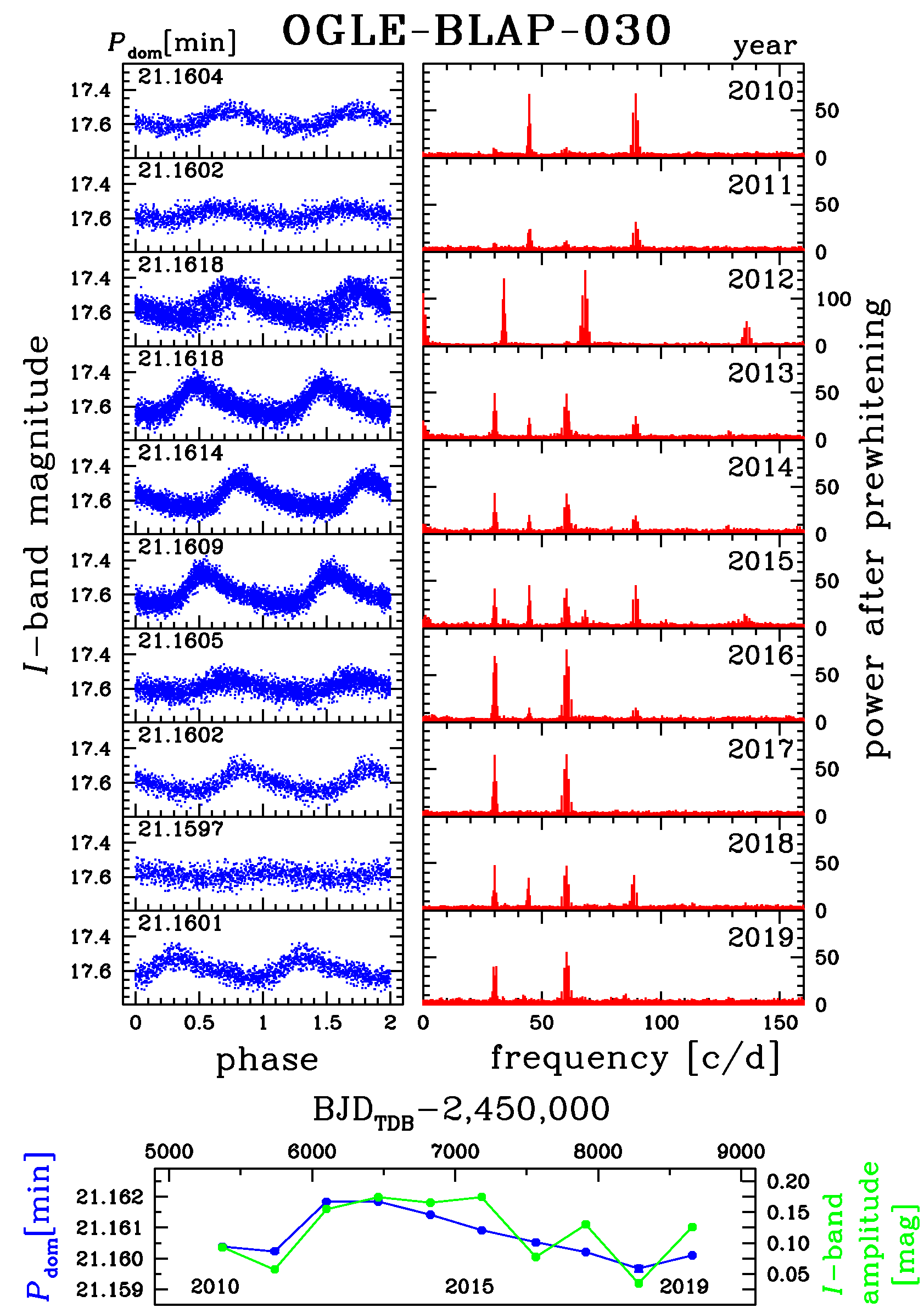}
\caption{Variations of the pulsation signal in the multimode object OGLE-BLAP-030. The OGLE-IV data are divided into seasons from 2010 to 2019. The left column shows, for each of the seasons, the $I$-band light curve phased with the dominant period. The magnitude range on the vertical axis is the same in each subpanel. The right column shows the power spectrum obtained after subtraction of the actual dominant period. Period and amplitude changes over the years are presented in the bottom panel. Period uncertainties are smaller than the dots.}
\label{fig:multimode30}
\end{figure}

\begin{figure}
\centering
\includegraphics[width=0.45\textwidth]{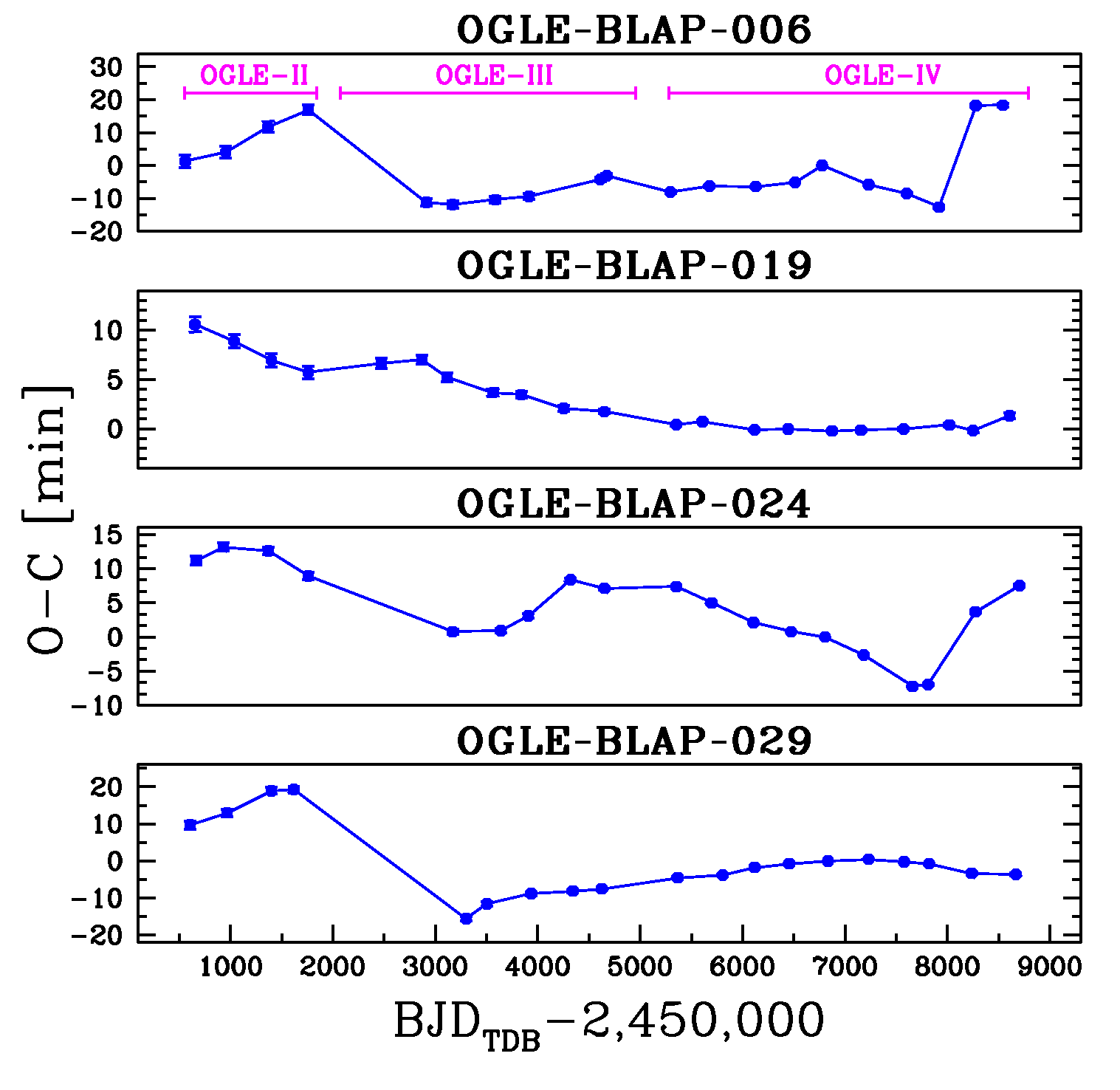}
\caption{O$-$C diagrams constructed for four BLAPs observed during three phases of the OGLE project, from OGLE-II to OGLE-IV covering years 1997--2019. The differences measured in OGLE-BLAP-019 indicate positive period changes in this pulsator.}
\label{fig:pdotO234}
\end{figure}

\begin{figure}
\centering
\includegraphics[width=0.45\textwidth]{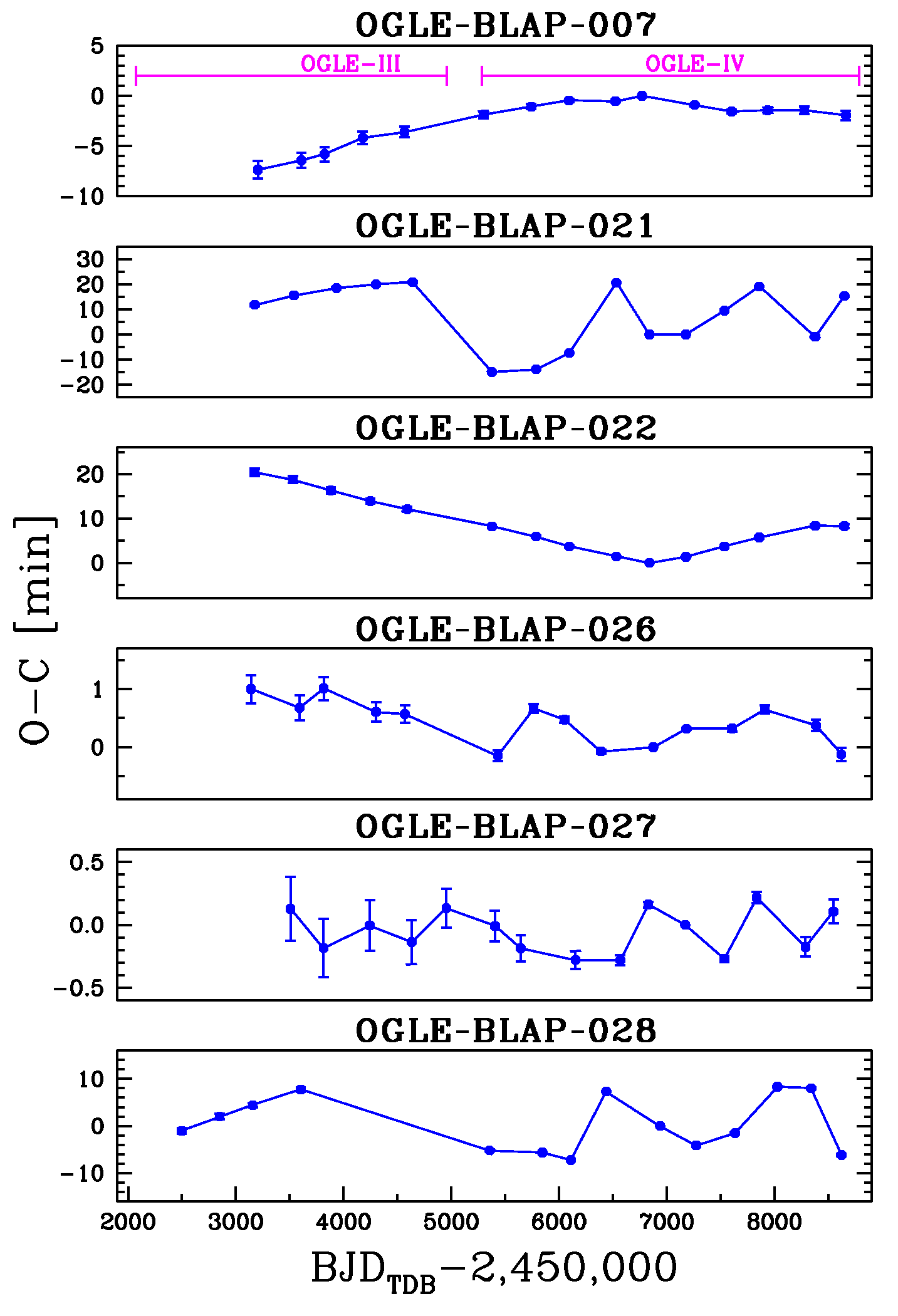}
\caption{O$-$C diagrams constructed for six selected BLAPs observed in OGLE-III and OGLE-IV over years 2001--2019. Objects OGLE-BLAP-007 and OGLE-BLAP-022 show evident negative and positive period changes, respectively. Changes in the remaining objects are irregular.}
\label{fig:pdotO34}
\end{figure}

\begin{figure}
\centering
\includegraphics[width=0.45\textwidth]{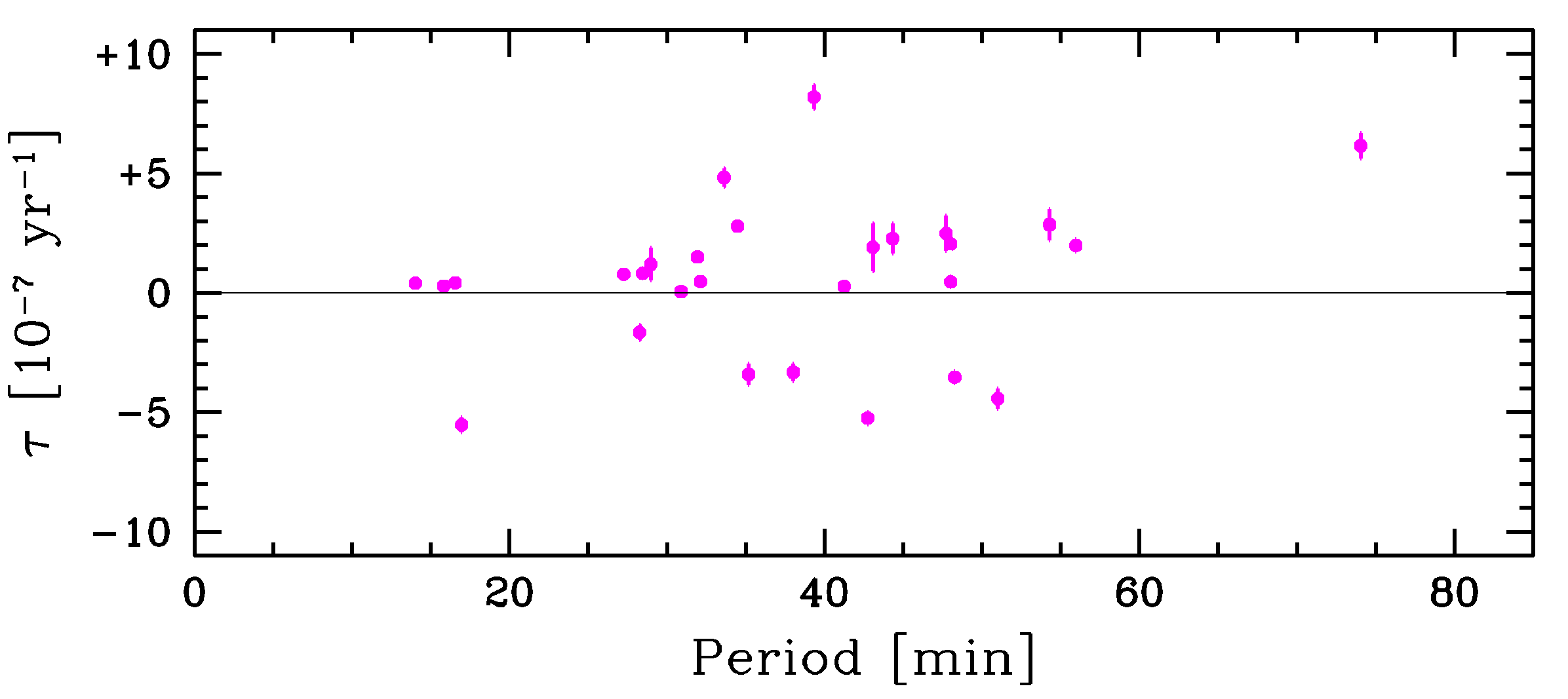}
\caption{Period change rates for BLAPs observed in OGLE-III and OGLE-IV. The rates were calculated based on the period values determined for each phase (Equation~(1)).}
\label{fig:pdot}
\end{figure}

Two BLAPs show more than one periodicity, namely objects OGLE-BLAP-001 and OGLE-BLAP-030. The former object, being the prototype of the whole class, was discovered in the OGLE-III data collected in years 2005--2006 \citep{2013AcA....63..379P}\footnote{At that time, the star was tentatively classifed as a $\delta$ Sct and it is identified as OGLE-GD-DSCT-0058 in \citet{2013AcA....63..379P} and \citet{2015AcA....65...63P}.}. The photometry revealed the presence of three periodic signals forming a symmetric triplet in the frequency space, with peaks at 49.06155(14), 50.96440(1), and 52.86723(13) cycles per day, separated from each other by about 1.9028 cycles per day. Hence, the dominant period is $P_{\rm dom}=28.255020(7)$ minutes, while the two additional periodicities are $P_1=27.238045(66)$ minutes and $P_2=29.350884(86)$ minutes. Such a triplet is present only in this object and is attributed to rotational splitting. OGLE-BLAP-001 was observed more recently in the fourth phase of OGLE (in years 2013--2020), but shallower and less frequent observations did not allow us to detect the side frequencies.

The second multiperiodic pulsator, OGLE-BLAP-030, shows strong amplitude and period changes. We analyzed the rich OGLE-IV photometry for this star by performing consecutive prewhitenings season by season from 2010 to 2019. The dominant signal is around 68.050 cycles per day, or at the period $P_{\rm dom}\approx21.161$ minutes. In Figure~\ref{fig:multimode30}, we present, for each season separately, a light curve folded with the actual dominant period and a power spectrum obtained after prewhitening with this value. We conclude that there are three independent frequencies in this star. Their exact values in 2015 were 68.04999(3), 89.31937(26), and 60.35173(26) cycles per day, or the periods $P_{\rm dom}=21.160914(9)$ minutes, $P_1=16.121923(47)$ minutes, and $P_2=23.860128(103)$ minutes, respectively. The remaining signals are subharmonics or combinations of the independent frequencies. The light-curve shape of the dominant signal is characteristic of pulsations in the fundamental radial mode. The ratios between the periods are the following: $P_1/P_{\rm dom}\approx0.762$, $P_{\rm dom}/P_2\approx0.887$, and $P_1/P_2\approx0.676$. The $P_1/P_{\rm dom}$ value is similar to the period ratio observed for stars pulsating in the fundamental mode and first overtone simultaneously, such as $\delta$ Sct-type stars \citep{2022MNRAS.510.1748N}. $P_1$ would then correspond to the first overtone. The source of the third periodicity ($P_2$) remains unknown. It could be a non-radial mode.

\begin{table*}
\centering \caption{Atmospheric Parameters and RVs Determined for the 15 BLAPs Observed with the MagE Spectrograph}
\begin{tabular}{ccccr}
\hline
Name          & $T_{\rm eff}$    & $\log g$        & $\log (N_{\rm He}/N_{\rm H})$ & $v_{\rm r}$ \\
              &       (K)        &                 &                               & (km s$^{-1}$) \\
\hline
OGLE-BLAP-001 & 30,800 $\pm 500$ & $4.61 \pm 0.07$ & $-0.55 \pm 0.05$ & $ 106\pm12$ \\
OGLE-BLAP-010 & 29,800 $\pm 300$ & $4.57 \pm 0.04$ & $-0.58 \pm 0.03$ & $ -48\pm12$ \\
OGLE-BLAP-019 & 28,000 $\pm 700$ & $4.29 \pm 0.09$ & $-0.76 \pm 0.08$ & $ 109\pm17$ \\
OGLE-BLAP-020 & 29,200 $\pm 500$ & $4.40 \pm 0.07$ & $-0.59 \pm 0.06$ & $ 174\pm28$ \\
OGLE-BLAP-021 & 28,500 $\pm 300$ & $4.46 \pm 0.04$ & $-0.64 \pm 0.03$ & $-210\pm9$ \\
OGLE-BLAP-022 & 28,900 $\pm 400$ & $4.45 \pm 0.06$ & $-0.74 \pm 0.05$ & $ 177\pm15$ \\
OGLE-BLAP-024 & 25,200 $\pm 300$ & $4.39 \pm 0.05$ & $-0.66 \pm 0.04$ & $-124\pm17$ \\
OGLE-BLAP-030 & 31,400 $\pm 300$ & $4.85 \pm 0.05$ & $-0.75 \pm 0.04$ & $ 139\pm13$ \\
OGLE-BLAP-031 & 26,800 $\pm 200$ & $4.38 \pm 0.03$ & $-0.54 \pm 0.03$ & $  85\pm12$ \\
OGLE-BLAP-033 & 33,100 $\pm 700$ & $5.04 \pm 0.11$ & $-0.88 \pm 0.07$ & $ 203\pm10$ \\
OGLE-BLAP-034 & 30,300 $\pm 300$ & $4.49 \pm 0.04$ & $-0.62 \pm 0.03$ & $ -68\pm16$ \\
OGLE-BLAP-037 & 32,800 $\pm 200$ & $4.93 \pm 0.04$ & $-2.15 \pm 0.05$ & $  -75\pm17$ \\
OGLE-BLAP-042 & 28,300 $\pm 1000$ & $4.19 \pm 0.14$ & $-0.52 \pm 0.12$ & $30\pm36$ \\
OGLE-BLAP-044 & 32,700 $\pm 200$ & $5.28 \pm 0.03$ & $-2.80 \pm 0.09$ & $ 21\pm12$ \\
OGLE-BLAP-049 & 29,300 $\pm 400$ & $4.91 \pm 0.06$ & $-0.67 \pm 0.04$ & $-11\pm15$ \\
\hline
\end{tabular}
\label{tab:spec}
\medskip
\end{table*}

During the decade of OGLE-IV observations of OGLE-BLAP-030, the $I$-band amplitude varied between 0.035 mag (in 2018) and 0.175 mag (in 2015) with about 0.002 mag error, while the dominant period changed between 21.159677(9) minutes (in 2018) and 21.161847(12) minutes (in 2013). We plot both the amplitude and dominant period variations in the bottom panel of Figure~\ref{fig:multimode30}. They are correlated in time. Uncertainties of the measured amplitudes are of 0.001~mag. In the case where the number of data points is lower than 300 the amplitude errors may reach 0.002 mag.

Five BLAPs, namely OGLE-BLAP-006, 019, 024, 029, and 030, have been observed over three OGLE phases, from OGLE-II to OGLE-IV. Figure~\ref{fig:pdotO234} illustrates the O$-$C diagrams constructed based on the times of maxima for the first four objects (only single-mode pulsators). The moment of maximum was taken to be the point closest to the maximum light in the phased light curve with at least 90 measurements in the season. A quasi-parabolic shape can be seen in the diagram for object OGLE-BLAP-019. This indicates a likely linear period change, here positive. For the remaining three stars, the changes are irregular. In the case of 23 BLAPs, the collected time-series data are shorter and cover only the OGLE-III and OGLE-IV phases. Figure~\ref{fig:pdotO34} illustrates the O-C diagrams for six such stars. In OGLE-BLAP-007, the period changes are negative, while in OGLE-BLAP-022, they are positive. Other stars exhibit irregular period variations.

We also calculated the period change rate $r$ in the most simple way as
\begin{equation}
r=\frac{\Delta P}{\Delta t}\frac{1}{P_{\rm IV}}
=\frac{P_{\rm IV}-P_{\rm III}}{t_{\rm IV}-t_{\rm III}}\frac{1}{P_{\rm IV}},
\end{equation}
where $t_{\rm III}$ and $t_{\rm IV}$ are the mean moments of the OGLE-III and OGLE-IV observations used to determine the periods $P_{\rm III}$ and $P_{\rm IV}$, respectively. This was computed for the prototype BLAP and 28 additional objects with OGLE-III and OGLE-IV data. Results are plotted in Figure~\ref{fig:pdot}. 

Amongst the 29 BLAPs, seven objects show negative period changes, while for the remaining 22 objects the changes are positive. In general, the pulsation periods in BLAPs change slowly, with a rate of several times $10^{-7}$ yr$^{-1}$. However, we note that the true period change rates may vary depending on the used sections of time. The outlying value in the case of OGLE-BLAP-030 likely stems from large variations on shorter timescales. There is no correlation between the rate of changes and the pulsation period.

\subsection{Atmospheric Parameters} \label{sec:atmo}

\begin{figure*}
\centering
\includegraphics[width=0.98\textwidth]{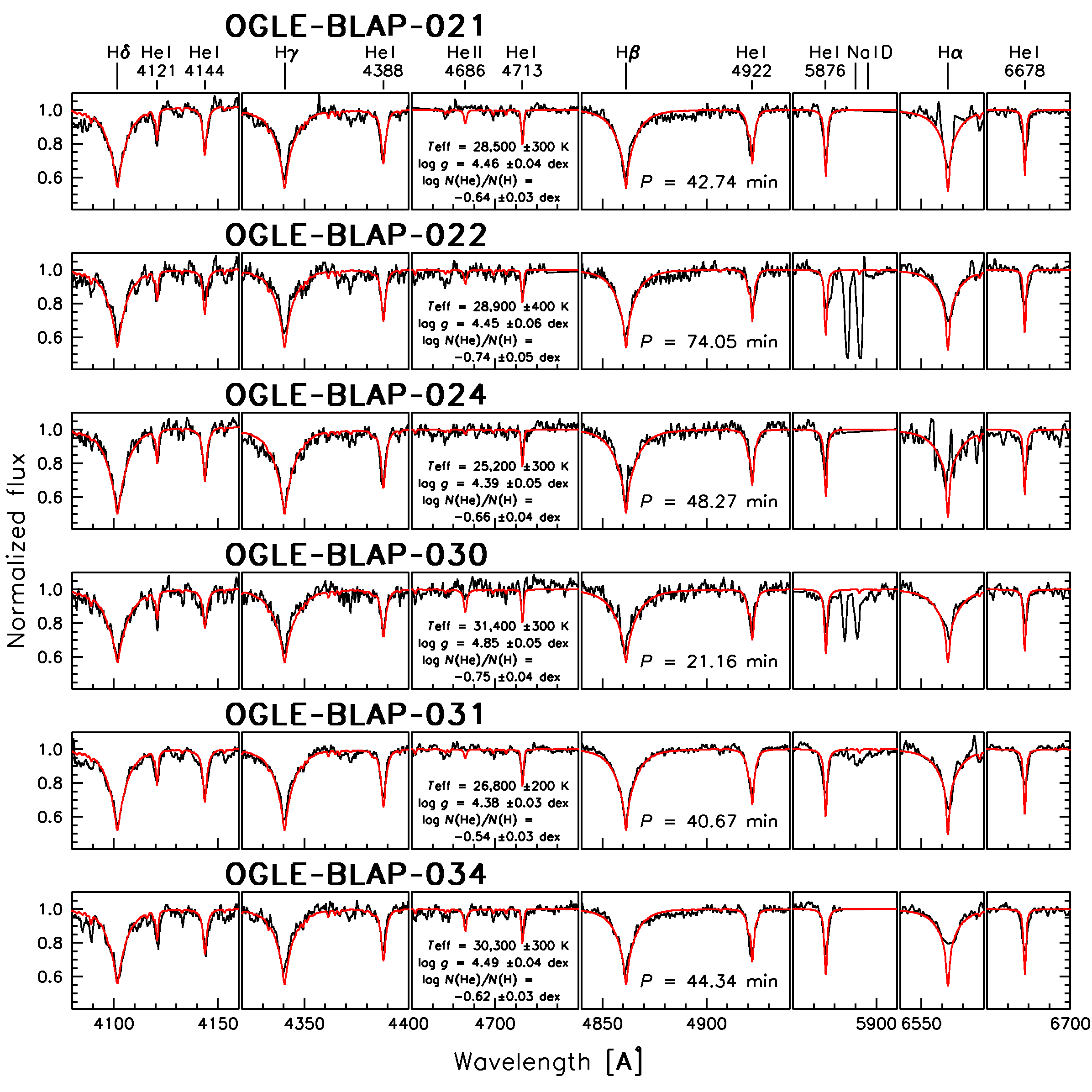}
\caption{Moderate-resolution spectra of six selected BLAPs. The best-fit synthetic spectra are shown with red lines. The interstellar Na~\textsc{i}~D lines are cut out for some of the spectra during the fitting procedure. The obtained atmospheric parameters as well as the pulsation period are provided for each object. Very similar parameters indicate that the stars form a homogeneous group of objects.}
\label{fig:newBLAPspec1}
\end{figure*}

\begin{figure*}
\centering
\includegraphics[width=0.98\textwidth]{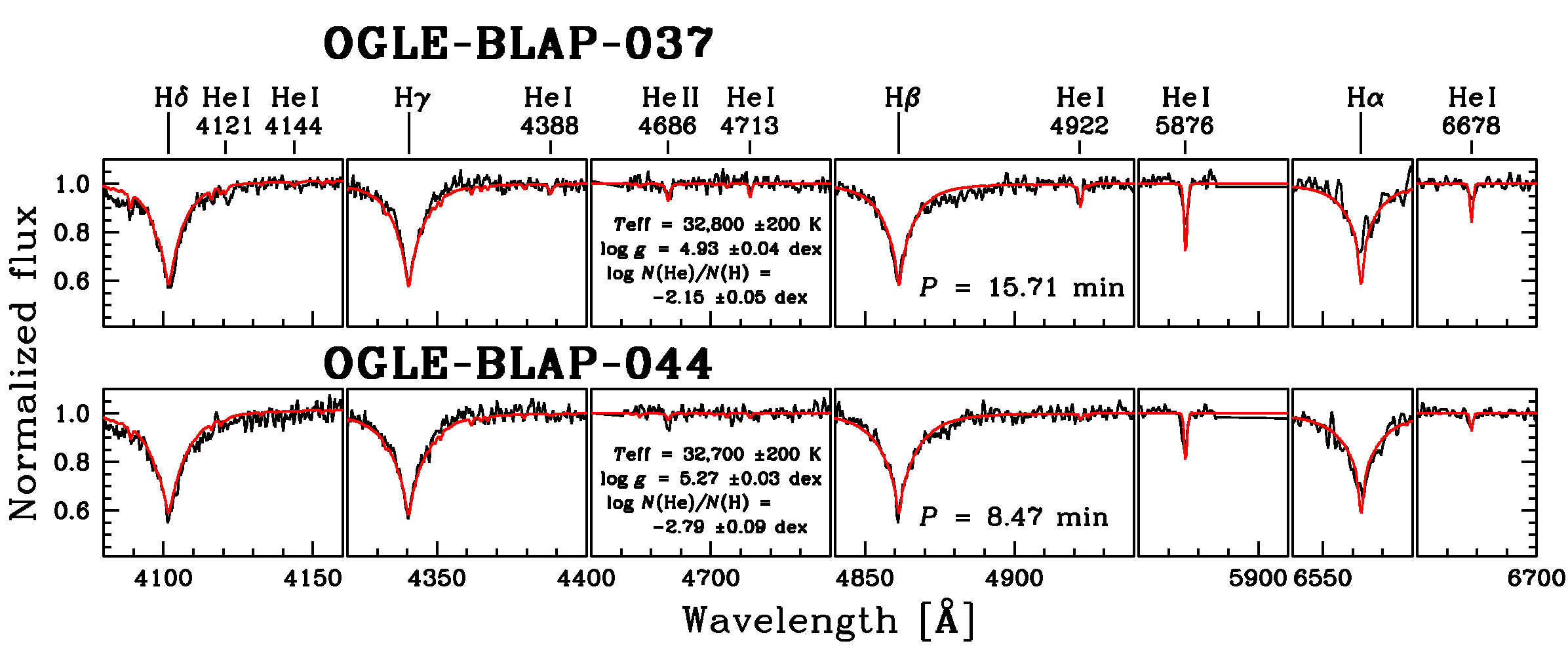}
\caption{Best fit to the spectra of two He-poor BLAPs in our spectroscopic sample: OGLE-BLAP-037 (= OW J1819-2729) and OGLE-BLAP-044 \citep{2023AcA....73....1B}.}
\label{fig:OmegaWhite4}
\end{figure*}

The atmospheric parameters of the 15 BLAPs observed with the MagE spectrograph are listed in Table~\ref{tab:spec} and our best fits for eight selected BLAPs are shown in Figs.~\ref{fig:newBLAPspec1}--\ref{fig:OmegaWhite4}. As explained in Section~\ref{sec:analysis}, the quoted uncertainties on the atmospheric parameters are the formal errors of the fitting procedure and the RV uncertainties are from the standard deviation of the RV values of single lines. The uncertainties do not include variations from the pulsations.

Among the 15 BLAPs observed with MagE, 13 stars have an atmosphere that is moderately enriched in helium. The helium abundance in these objects varies within \nhe\ $=-0.9$ to $-0.5$ dex. They have effective temperatures between 25,000~K and 34,000~K and surface gravities between $\log g = 4.3$ and 5.0. The two remaining stars, OGLE-BLAP-037 and OGLE-BLAP-044, have an atmosphere depleted in helium, with \nhe\ $< -2$ dex (see Figure~\ref{fig:OmegaWhite4}). In fact, with a pulsation period of only 8.47~minutes, OGLE-BLAP-044 has properties similar to the known high-gravity (and also He-poor) BLAPs that have periods between 3 and 8~minutes. On the other hand, OGLE-BLAP-037 has a significantly longer period (15.71~minutes). Incidentally, the variable nature of this object was reported by \citet{2022MNRAS.513.2215R} who identified it in the OmegaWhite survey and referred to it as OW~J1819-2729. However, the authors classified it as a much cooler object (with \teff=12,000~K) based on follow-up spectroscopy with a lower resolving power ($R\sim2000$). Our spectrum, shown in Figure~\ref{fig:OmegaWhite4}, indicates a high \teff, notably due to the presence of the He~\textsc{ii} $\lambda$4686 line. In addition, the RV of the Ca~\textsc{ii}~K line visible in the spectrum is not consistent with the stellar RV, indicating that the Ca line does not originate from the stellar photosphere.


\begin{figure*}
\centering
\includegraphics[width=0.45\textwidth]{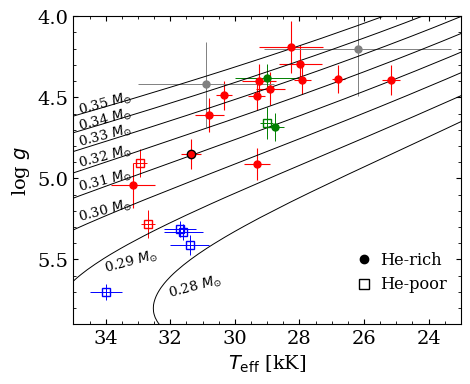}
\includegraphics[width=0.45\textwidth]{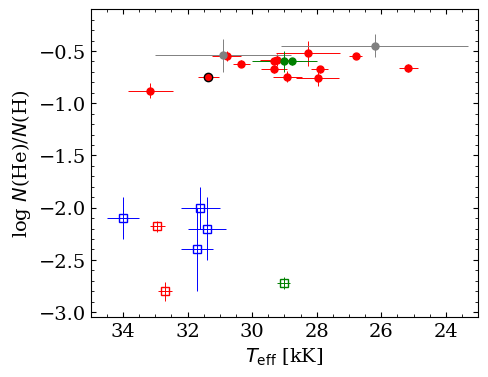}
\caption{Location of the BLAPs in the atmospheric parameter spaces. The 15 stars included in this work are indicated with red symbols and OGLE-BLAP-030 (multimode pulsator) is indicated with an additional black circle. The four high-gravity BLAPs of \citet{2019ApJ...878L..35K} are plotted in blue. The grey symbols are for the two BLAPs with low-resolution spectra analyzed in \citet{2017NatAs...1E.166P}, and the green symbols are for TMTS-BLAP-1, OGLE-BLAP-009, and SMSS-BLAP-1 \citep{2023NatAs...7..223L,2024MNRAS.52710239B,2024MNRAS.529.1414C}. Open squares indicate BLAPs with a He-poor atmosphere and filled circles BLAPs with a He-rich atmosphere. The evolutionary tracks for low-mass He-core pre-WDs are from \citet{2019ApJ...878L..35K}. }
\label{fig:param}
\end{figure*}

Figure~\ref{fig:param} presents the atmospheric parameters of BLAPs in the \teff--$\log g$ and \teff--helium abundance planes. In this figure, we include the 15 BLAPs observed with MagE, two BLAPs for which low-resolution spectra were analyzed in \citet{2017NatAs...1E.166P}, the four high-gravity BLAPs from \citet{2019ApJ...878L..35K}, and TMTS-BLAP-1, SMSS-BLAP-1, and OGLE-BLAP-009 \citep{2023NatAs...7..223L,2024MNRAS.529.1414C,2024MNRAS.52710239B}. In the \teff--$\log g$ diagram (left panel in Figure~\ref{fig:param}), the BLAPs loosely follow a sequence where the surface gravity increases with the effective temperature. If they are He-core pre-WDs, they would have masses between 0.3 and 0.35~\Msun\ according to the evolutionary sequences of \citet{2019ApJ...878L..35K}.  The four high-gravity BLAPs appear to follow a similar sequence but shifted to larger gravities.

The second diagram presents the \teff--helium abundance plane (right panel in Figure~\ref{fig:param}), where the BLAPs split up into two distinct groups: those, like the prototype, having a He-rich atmosphere, and those, like the high-gravity BLAPs, having a He-poor atmosphere. Although both groups overlap in \teff, the He-poor stars are mainly found at the hot end (\teff\ $>$ 31,000~K) of the temperature range covered by the BLAPs (i.e., 25,000--34,000~K). 

\subsection{Metallicity} \label{sec:abun}

\begin{figure*}
\centering
\includegraphics[angle=90, width=0.98\textwidth]{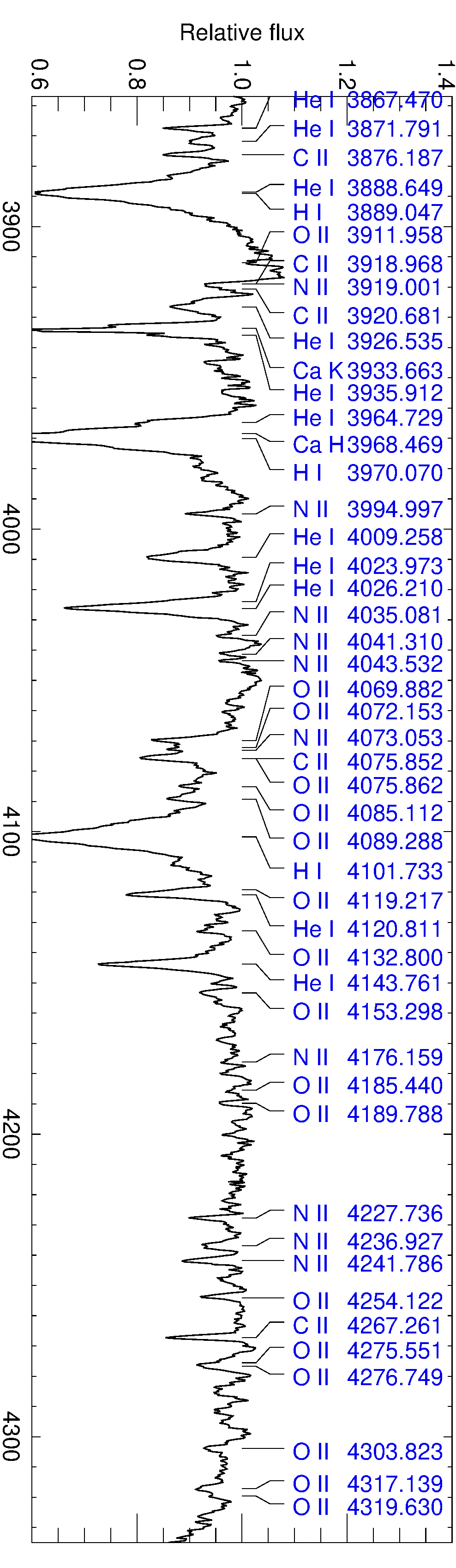}
\includegraphics[angle=90, width=0.98\textwidth]{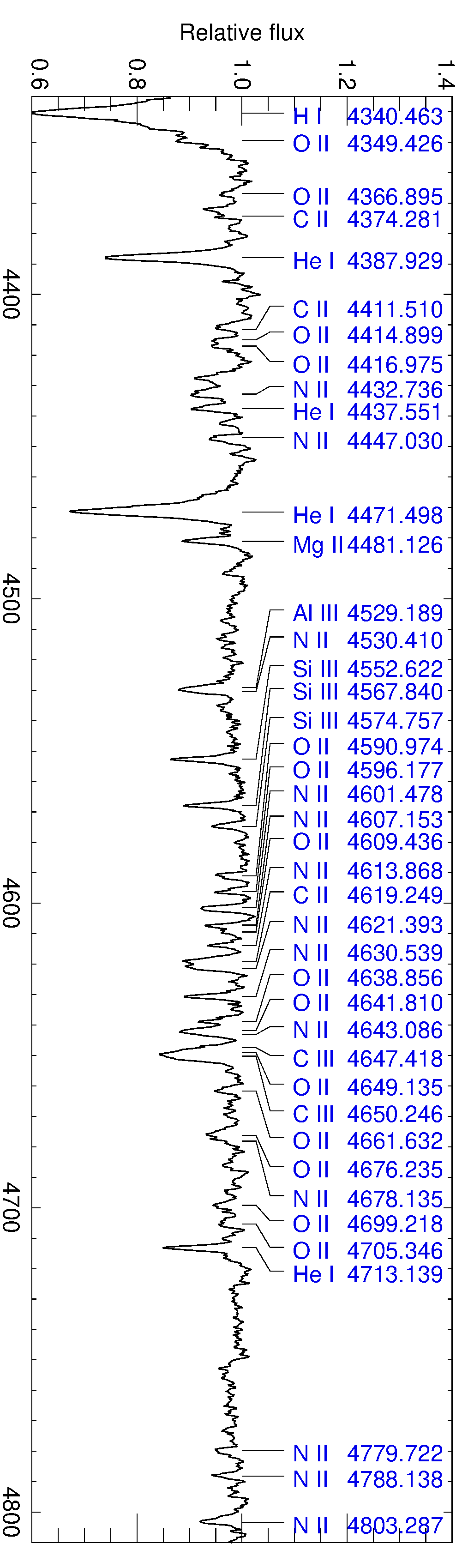}
\includegraphics[angle=90, width=0.98\textwidth]{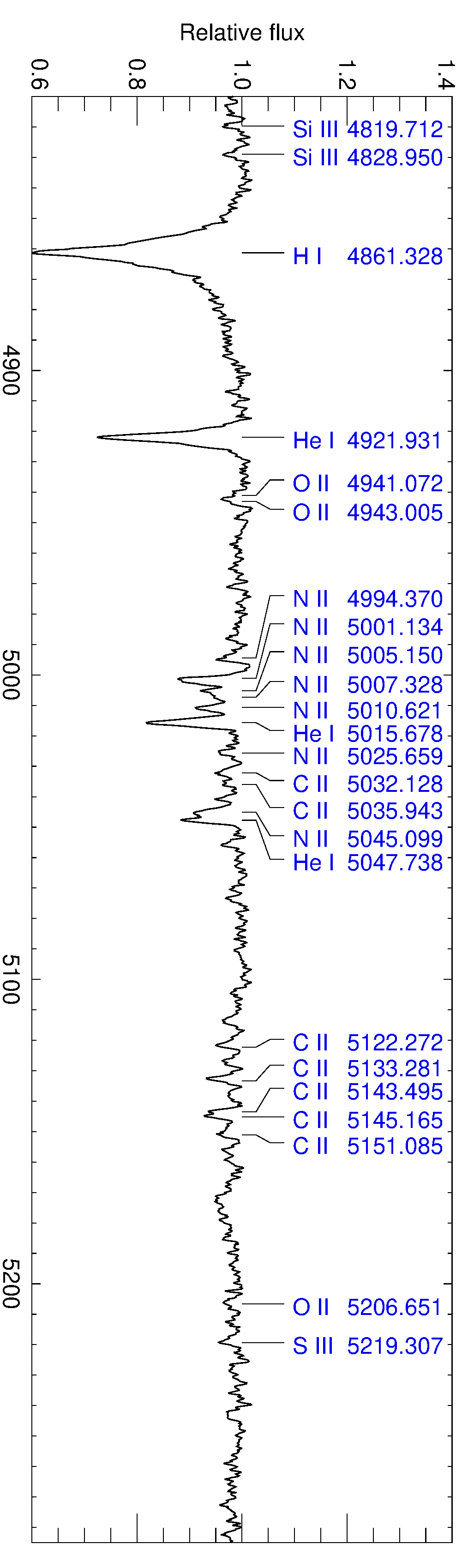}
\includegraphics[angle=90, width=0.98\textwidth]{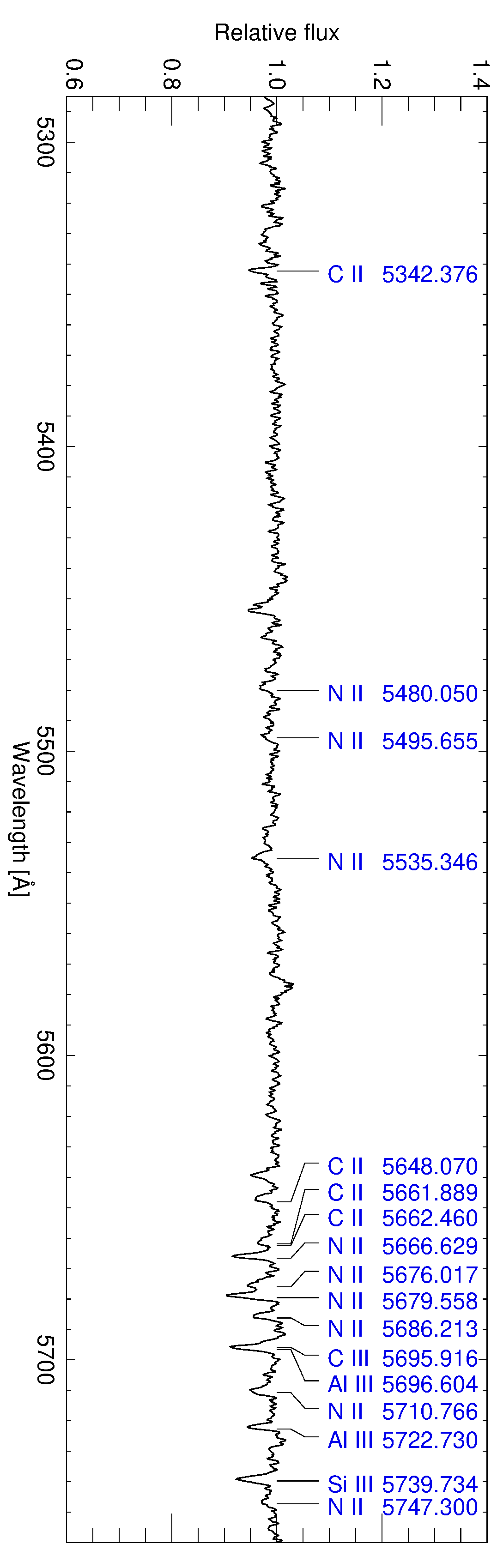}
\caption{Spectroscopic lines identified in He-rich BLAPs. The spectrum is a combination of the RV-corrected spectra of OGLE-BLAP-010, 019, 021, 031, and 034. The ion and wavelength (in angstroms) of the strongest transitions are marked.}
\label{fig:elements}
\end{figure*}

Besides the hydrogen and helium lines, many additional absorption features are visible in the spectra of the He-rich BLAPs. The majority of these spectral lines are present with a similar strength in the spectra of all He-rich BLAPs (at least in those having a sufficiently high signal-to-noise ratio (S/N)), indicating that they share a similar atmospheric composition. To identify metal lines, we combined the RV-corrected spectra of five BLAPs having good S/N and \teff\ between 25,000 and 30,000 K, namely OGLE-BLAP-010, 019, 021, 031, and 034. The resulting spectrum is shown in Figure~\ref{fig:elements}. We were able to identify all prominent features. The majority of the metal lines originate from C~\textsc{ii}, C~\textsc{iii}, N~\textsc{ii}, and O~\textsc{ii}. A few additional features are from Mg~\textsc{ii}, Al~\textsc{iii}, and Si~\textsc{iii}.

\begin{figure*}
\centering
\includegraphics[angle=90,width=0.48\textwidth]{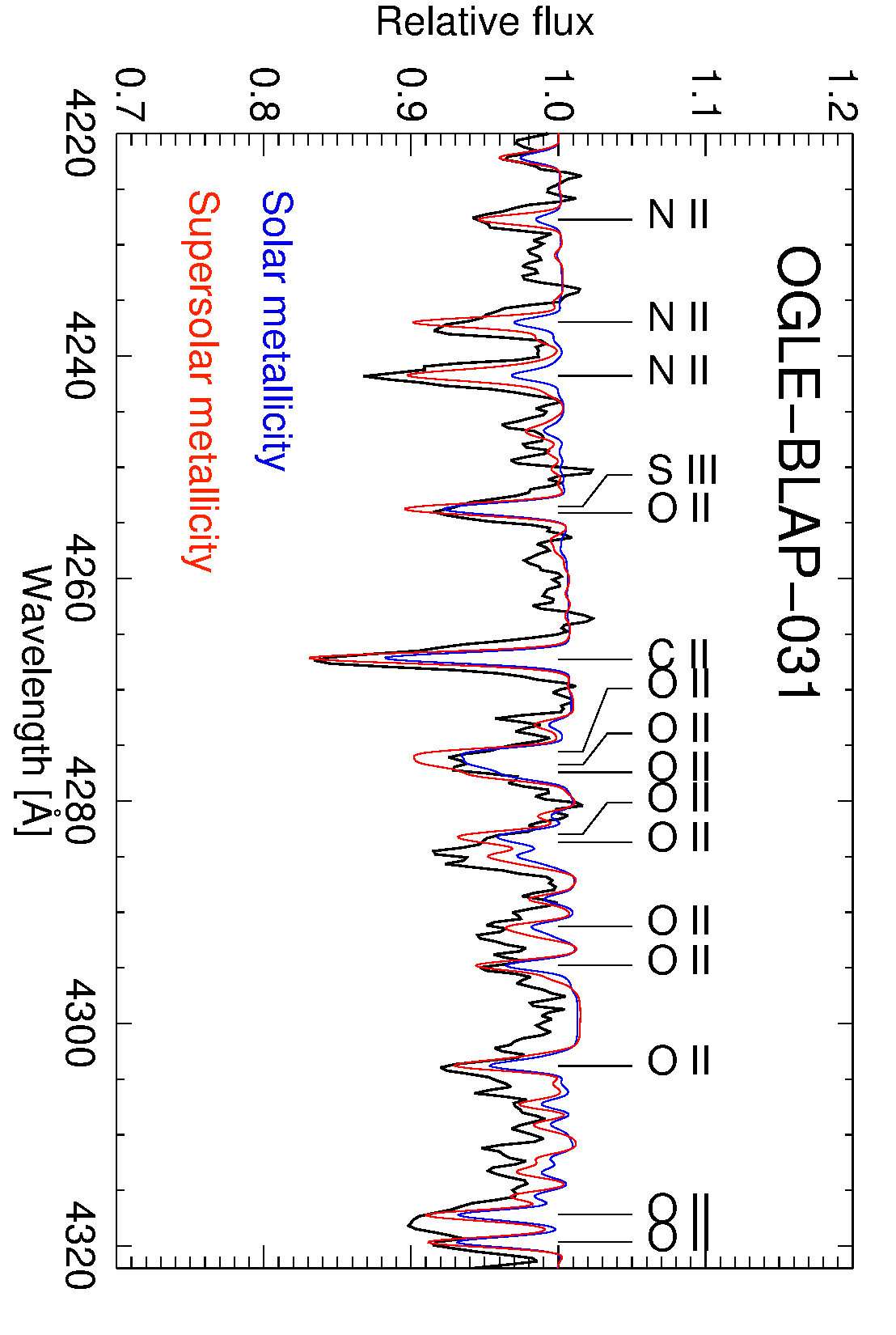}
\includegraphics[angle=90,width=0.48\textwidth]{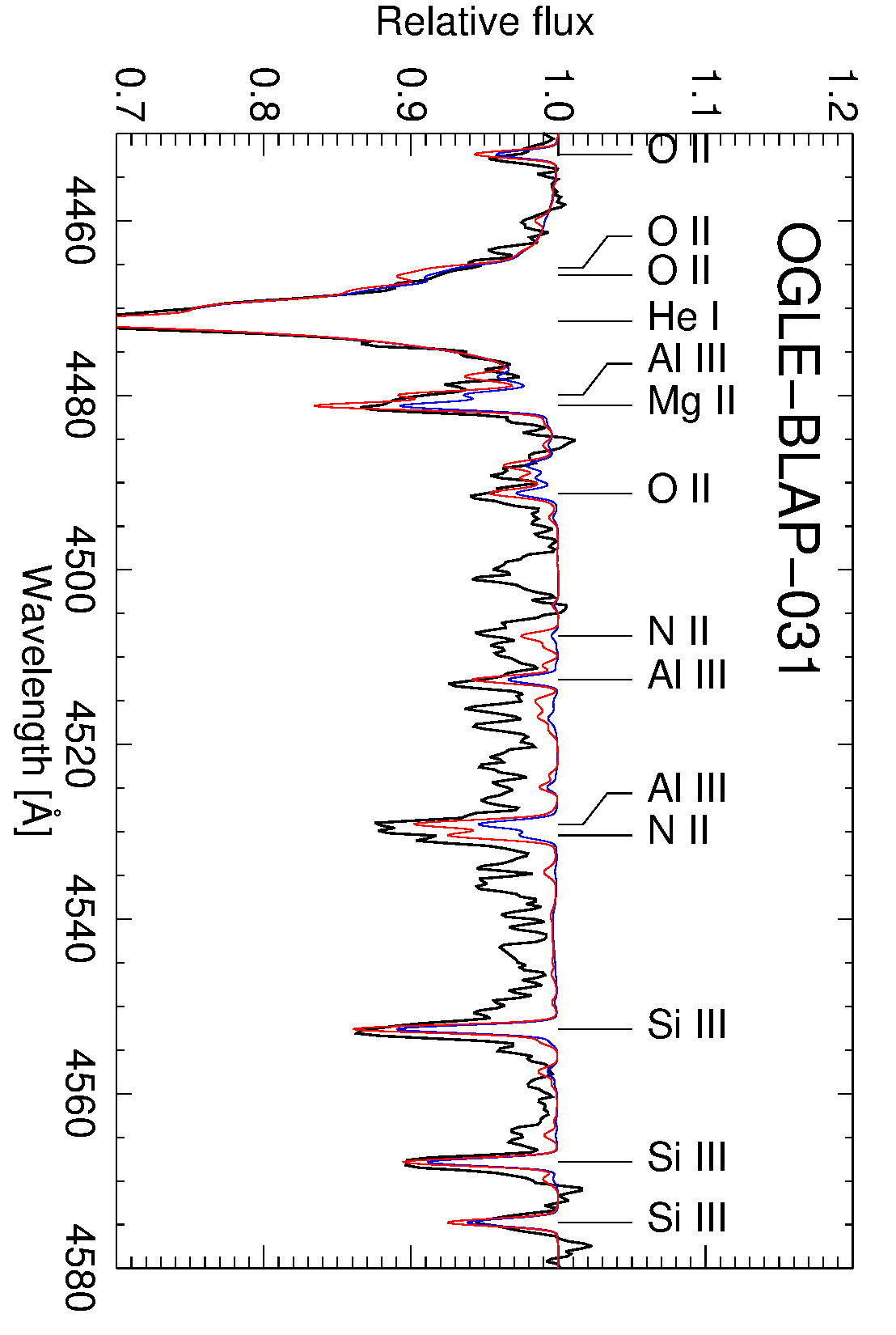}
\includegraphics[angle=90,width=0.9\textwidth]{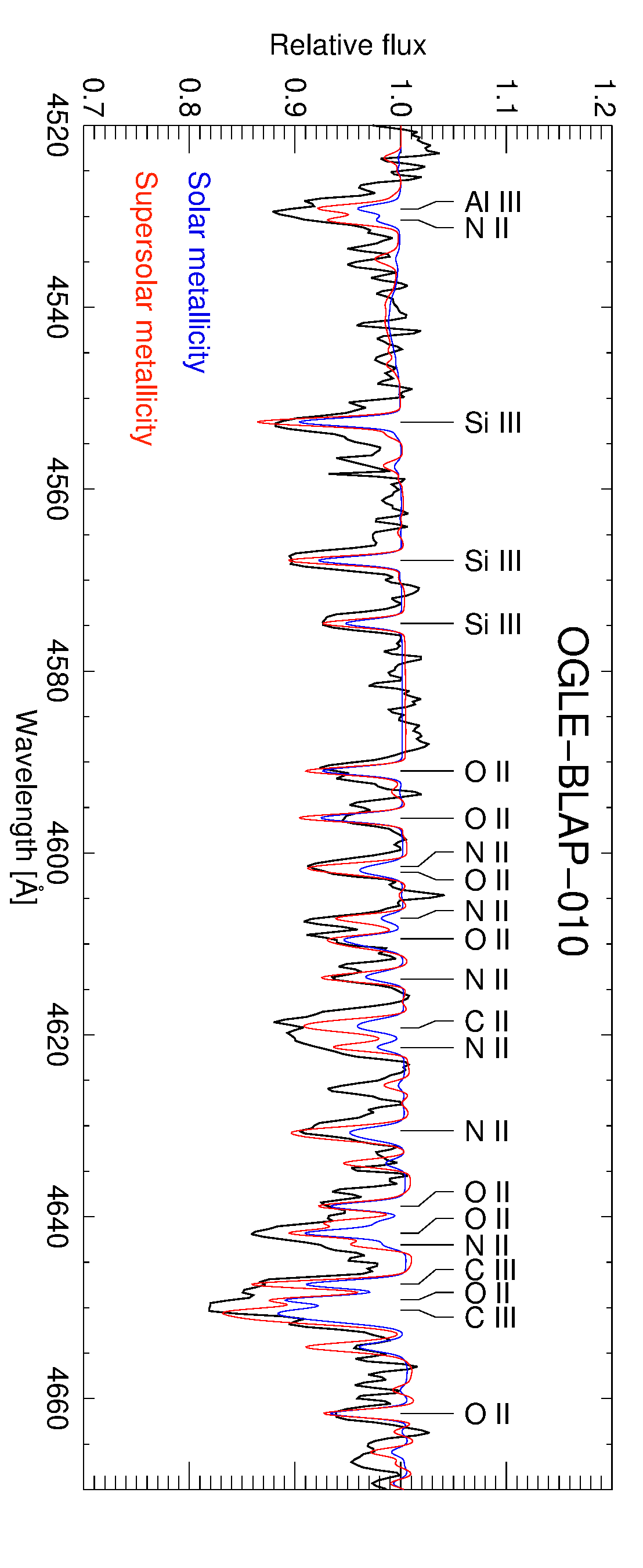}
\caption{Synthetic spectra with solar (blue line) and supersolar abundances (red line) compared to the spectra of OGLE-BLAP-031 (top) and OGLE-BLAP-010 (bottom). }
\label{fig:models}
\end{figure*}

To provide an estimate of the elemental abundances of the He-rich BLAPs, we computed model atmospheres and synthetic spectra with varying abundances for two selected stars with good S/N, OGLE-BLAP-010 and OGLE-BLAP-031. We first computed model atmospheres and spectra having the specific parameters of each star and included a solar abundance composition. The resulting spectra, compared to the observations, are displayed in Figure~\ref{fig:models} for three selected wavelength ranges. A solar composition results in many lines being weaker than in the observations, notably those of nitrogen. We proceeded into incrementally increasing the individual abundances of the elements mentioned above. For these two stars, we estimated the following abundances: carbon, magnesium, and aluminum $\sim$5$\times$ solar; nitrogen $\sim$10$\times$ solar; and oxygen $\sim$2$\times$ solar. Silicon appeared to be enhanced by up to 5$\times$ solar in OGLE-BLAP-010 but only up to 2$\times$ solar in OGLE-BLAP-031.

It is not uncommon for evolved stars with an atmosphere enriched in helium to also show enhancement in nitrogen and/or carbon. For example, the He-rich subluminous O and B stars are typically enriched in either N or C, and sometimes both \citep{2010MNRAS.409..582N,2007A&A...462..269S,2009JPhCS.172a2015H}. The BLAPs seem to share these characteristics as well, with an enhancement in nitrogen more pronounced than that in carbon.

\subsection{Period-Gravity Relationship} \label{sec:pg}

\begin{figure}
\centering
\includegraphics[width=0.48\textwidth]{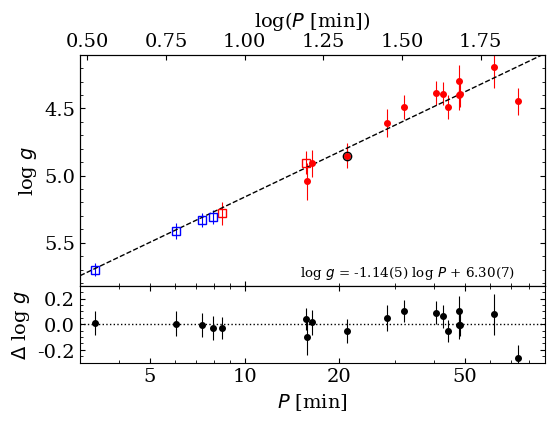}
\caption{Period--gravity diagram for the BLAPs. The 15 BLAPs observed with MagE are plotted in red and the four high-gravity BLAPs of \citet{2019ApJ...878L..35K} are plotted in blue. Open squares indicate He-poor BLAPs and filled circles He-rich BLAPs. OGLE-BLAP-030 (multimode pulsator) is indicated with an additional black circle. The dashed line indicates our linear fit with the equation. The bottom panel shows residuals from the best fit in terms of $\log g$.}
\label{fig:pg}
\end{figure}

The final parameter space we explore is the period--gravity plane. The period $P$ of a radial fundamental mode is determined by the mean stellar density $\bar{\rho}$ so that:
\begin{equation}
P \propto \left (\bar{\rho} \right )^{-1/2} \propto \left ( \frac{M}{R{}^3} \right )^{-1/2},
\end{equation}
where $M$ is the mass of the star and $R$ its radius. Since the gravity $g \propto M/R^2$, after transformations we obtain
\begin{equation}
P \propto {M}^{1/4} {g}^{-3/4}.
\end{equation}
In Figure~\ref{fig:pg}, we plot the surface gravity and period on logarithmic scales and find that they follow a linear relationship as expected from Equation~(3). We fitted a linear regression in the $\log P$ versus $\log g$ plane presented in Figure~\ref{fig:pg} using the 15 stars from this paper (the ones listed in Table~\ref{tab:spec}), and the four high-gravity BLAPs of \citet{2019ApJ...878L..35K}. For the He-rich BLAPs, we restricted ourselves to the BLAPs observed with the MagE spectrograph because they form a homogeneous sample. We fitted the following equation:
\begin{equation}
\log g = a \cdot \log P + b
\end{equation}
where the slope is $a=-1.14\pm0.05$, the intercept is $b=6.30\pm0.07$, and where $P$ is expressed in minutes. The bottom panel of Figure~\ref{fig:pg} shows the residuals in log~$g$ between the measured value and that predicted by the regression. Only the longest-period object, OGLE-BLAP-022, deviates significantly from the relation, that is by about $4\sigma$. The mean residual, in absolute value, when removing this star, is 0.05 dex in $\log g$, which is very reasonable given the limited accuracy of the surface gravity measurements.

An interesting feature visible in the period--gravity plane of Figure~\ref{fig:pg} is the separation, in terms of period length, between the He-poor BLAPs at short periods and the He-rich objects having longer periods. This separation happens between the He-rich star OGLE-BLAP-033 (P=15.82 minutes) and the He-poor star OGLE-BLAP-037 (P=15.71 minutes). The two stars have very close periods, and a similar location in the \teff--$\log g$ diagram, but OGLE-BLAP-037 has a He-poor atmosphere like that of the high-gravity BLAPs. It is also worth noting that among the four BLAPs identified by \citet{2022MNRAS.513.2215R}, OW-BLAP-1 has a period of 10.85 minutes and a He-poor atmosphere, while the other three objects have $P>20$ minutes and He-rich atmospheres\footnote{They are not included in Figure~\ref{fig:pg}.}. However, the latest addition to the BLAPs family, SMSS-BLAP-1, is a He-poor object with a period of 19.52 minutes \citep{2024MNRAS.529.1414C}.

\subsection{P-L Relation} \label{sec:pl}

Since the luminosity $L$ of the star is inversely proportional to its surface gravity $g$,
\begin{equation}
L \propto {R}^2 {T_{\rm eff}}^4 \propto \frac {M {T_{\rm eff}}^4}{g},
\end{equation}
and assuming that the masses and effective temperatures are similar, we expect that radially pulsating BLAPs obey a linear $P$-$L$ relation like classical pulsators---Cepheids, for instance. The $P$-$L$ relation can be written in the form
\begin{equation}
M_{\rm bol} = a' \cdot \log P + b',
\end{equation}
where $M_{\rm bol}$ is the bolometric absolute brightness of the star. To find the coefficients of the relation, $a'$ and $b'$, we can use the information on the brightest BLAP, OGLE-BLAP-009, recently investigated in \cite{2024MNRAS.52710239B}. The star has an average effective temperature of $T_{\rm eff}=28,000\pm1500$ K and surface gravity of $\log g=4.40\pm0.20$ dex. The estimated parallax is $\varpi=0.46\pm0.04$ mas after zero-point correction \citep{2021A&A...649A...4L} of the value provided in Gaia EDR3 \citep{2021A&A...649A...1G}. This gives a geometric distance of $d=2174$~pc with a $1\sigma$ deviation at $d_{\rm low}=2000$~pc and $d_{\rm high}=2381$~pc. At such distance and Galactic latitude of $b=-1\fdg66$, OGLE-BLAP-009 is located in moderately obscured regions of the thin Galactic disk. According to stellar models prepared by \cite{2013ApJS..208....9P}, such a hot object has the intrinsic color of $(V-I)_0=-0.33$ mag and bolometric correction ${\rm BC}_V=-2.75$ mag. The observed mean $V-I$ color of OGLE-BLAP-009 is $15.64-15.07=+0.57$ mag, which implies a reddening of $E(V-I)=(V-I)-(V-I)_0=0.90$ mag. We can calculate the $I$-band extinction to the star as $A_I=R_{I,VI} \cdot E(V-I)=1.08\pm0.13$ mag, where $R_{I,VI}=1.20\pm0.14$ is the total-to-selective extinction ratio in this direction derived from the reddening map prepared by \cite{2013ApJ...769...88N} based on OGLE-III observations of bulge red clump stars. Hence, we get the unreddened brightness values of $I_0=I-A_I=13.96\pm0.13$ mag and $V_0=I_0+(V-I)_0=13.60\pm0.13$ mag, and further the absolute magnitude in $V$, $M_V=V_0-5 \cdot \log d+5=+1.97^{+0.31}_{-0.33}$ mag, and the bolometric brightness of the star $M_{\rm bol}=M_V+{\rm BC}_V=-0.78^{+0.31}_{-0.33}$ mag. Assuming the bolometric brightness of the Sun $M_{\rm bol,\odot} = +4.74$ mag, we find the luminosity of OGLE-BLAP-009 in solar units as $\log(L/L_{\rm \odot})=-0.4(M_{\rm bol}-M_{\rm bol,\odot})=2.21^{+0.13}_{-0.12}$ or $L=161^{+58}_{-40}L_{\rm \odot}$. This result is in perfect agreement with the value of $L=170^{+60}_{-50}L_{\rm \odot}$ obtained through the fitting of spectral energy distribution in \cite{2024MNRAS.52710239B}. Finally, we calculate the slope of the $P$-$L$ relation as $a'=2.5 \cdot a$ and the intercept as $b'=M_{\rm bol} + 2.5 \cdot a \cdot \log P$ mag for the pulsation period $P=31.94$ minutes. The obtained $P$-$L$ relation with uncertainties is as follows:
\begin{equation}
M_{\rm bol} = (-2.85\pm0.12) \cdot \log (P~[{\rm minutes}]) + (3.5\pm0.5).
\end{equation}

\subsection{Kinematics and Location of the BLAPs in the Milky Way} \label{sec:kinem}

\begin{figure}
\centering
\includegraphics[width=0.45\textwidth]{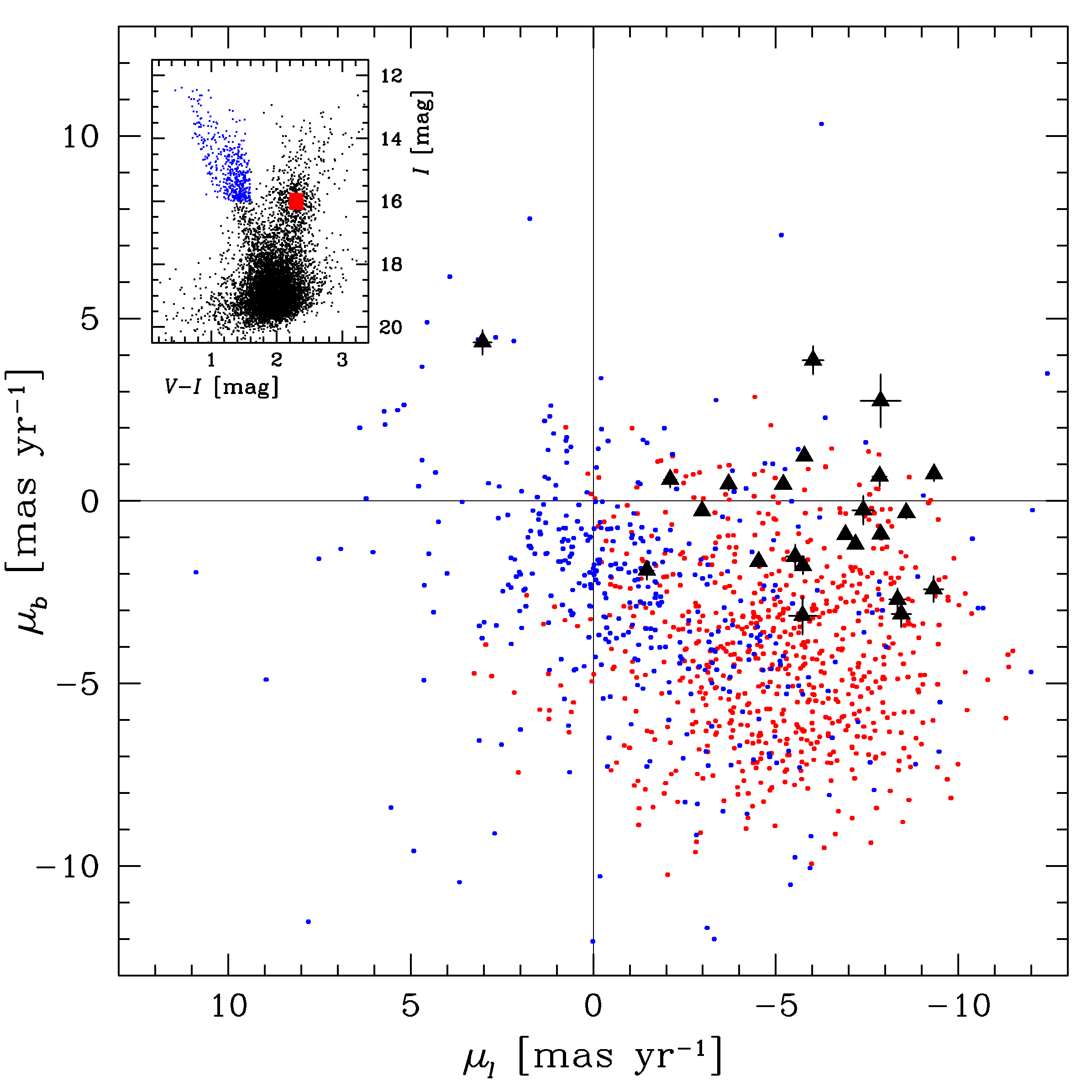}
\caption{Proper motions in Galactic coordinates of 23 BLAPs with reliable astrometric solution (black triangles) compared to main sequence stars from the Milky Way's disk (blue points) and red clump giants from the Milky Way's bulge (red points) selected from the color-magnitude diagram in the inset box. The observed sample of BLAPs is a mixture of objects located in the Galactic bulge and Galactic disk with the majority likely from the bulge.}
\label{fig:pm}
\end{figure}

In Table ~\ref{tab:gaia}, we list parallaxes and proper motions derived from Gaia EDR3 \citep{2021A&A...649A...1G} for 23 BLAPs observed toward the Galactic bulge with reliable astrometric solution (the renormalised unit weight error RUWE $<1.4$)\footnote{The given parallaxes are not corrected for the zero point.}. The proper motions were transformed from the equatorial to the galactic system using equations given in \cite{2013arXiv1306.2945P}. The OGLE BLAPs are distant objects located at several kiloparsecs from the Sun. Unfortunately, their parallaxes have large uncertainties. Only the brightest object in the sample, OGLE-BLAP-009, has a parallax determined better than $5\sigma$. In Figure~\ref{fig:pm}, we overlay the positions of the BLAPs in the proper-motion plane onto the positions of red clump giants from the Galactic bulge and foreground main-sequence stars from the Galactic disk. The selected stars are indicated in the attached color-magnitude diagram constructed for a $9'\times18'$ OGLE-IV subfield at ($l$,$b$)=($0\degr$,$-4\degr$). The disk stars are not completely separated from the bulge stars in the proper-motion plane, but the BLAPs group mainly around the area occupied by the latter objects. Certainly, the following two objects are in the disk: OGLE-BLAP-009 (based on the parallax) and OGLE-BLAP-013 (based on the proper motion).

\begin{table}[t!]
\centering \caption{Parallaxes and Proper Motions of 23 BLAPs toward the Galactic Bulge with a Reliable Astrometric Solution Based on Gaia EDR3}
\begin{tabular}{cccc}
\hline
Name          & $\varpi$ &      $\mu_l$    & $\mu_b$ \\
              &  (mas)   & (mas yr$^{-1}$) & (mas yr$^{-1}$) \\
\hline
OGLE-BLAP-002 & 0.30(22) & -5.53(24) & -1.52(33) \\
OGLE-BLAP-003 & 0.03(27) & -8.44(27) & -3.10(36) \\
OGLE-BLAP-004 & 0.02(20) & -2.10(16) &  0.59(22) \\
OGLE-BLAP-005 & 2.03(93) & -7.87(55) &  2.74(73) \\
OGLE-BLAP-006 & 0.21(24) & -9.32(28) & -2.41(36) \\
OGLE-BLAP-009 & 0.41(4)  & -5.20(4)  &  0.45(5)  \\
OGLE-BLAP-010 & 0.21(10) & -6.91(8)  & -0.92(11) \\
OGLE-BLAP-011 & 0.00(11) & -5.78(9)  &  1.23(13) \\
OGLE-BLAP-013 & 0.52(29) &  3.04(25) &  4.35(34) \\
OGLE-BLAP-014 & 0.54(8)  & -7.88(6)  & -0.91(9)  \\
OGLE-BLAP-016 & 1.12(23) & -7.84(20) &  0.67(27) \\
OGLE-BLAP-020 & 0.00(14) & -9.34(14) &  0.74(19) \\
OGLE-BLAP-021 & 0.26(15) & -8.58(12) & -0.32(16) \\
OGLE-BLAP-023 & 0.38(22) & -5.74(17) & -1.78(22) \\
OGLE-BLAP-026 & 0.32(24) & -8.34(23) & -2.70(31) \\
OGLE-BLAP-027 & 0.34(27) & -6.02(28) &  3.85(39) \\
OGLE-BLAP-029 & 0.34(16) & -3.71(13) &  0.47(17) \\
OGLE-BLAP-031 & 0.09(8)  & -7.18(7)  & -1.18(10) \\
OGLE-BLAP-032 & 0.06(22) & -1.47(19) & -1.90(25) \\
OGLE-BLAP-033 & 0.26(59) & -5.73(38) & -3.14(52) \\
OGLE-BLAP-034 & 0.54(14) & -4.53(12) & -1.66(16) \\
OGLE-BLAP-035 & 0.97(38) & -7.40(30) & -0.25(39) \\
OGLE-BLAP-036 & 0.25(6)  & -2.98(6)  & -0.27(9)  \\
\hline
\end{tabular}
\label{tab:gaia}
\medskip
\end{table}

\section{Discussion} \label{sec:disc}

\subsection{Period-Gravity Relationship: Comparison with Models}

\begin{figure}
\centering
\includegraphics[width=0.48\textwidth]{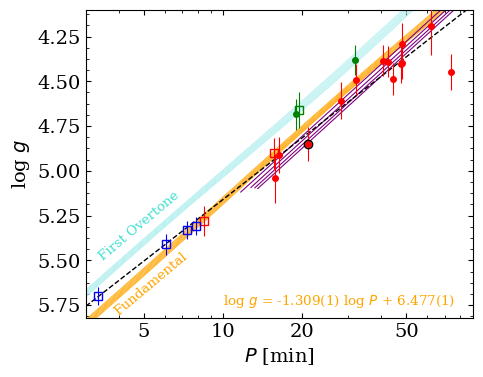}
\caption{Period--gravity diagram for BLAPs. The symbols are as described in Figure~\ref{fig:param}, with the addition of TMTS-BLAP-1, OGLE-BLAP-009, and SMSS-BLAP-1 in green. The dashed line is our fitted relationship to the observations. The orange (turquoise) stripe represents the predicted periods for the fundamental (first overtone) of the low-mass pre-WDs evolutionary models shown in Figure~\ref{fig:param} \citep{2019ApJ...878L..35K}. The equation describes the log~$P-$log~$g$ relation for the fundamental mode of these evolutionary models. The purple lines show the predicted periods for the fundamental modes of He-shell burning stars with masses of 0.5--0.8~\Msun \citep{2023NatAs...7..223L}.}
\label{fig:pg_models}
\end{figure}

The $P$-$L$ (or period--gravity) relationship obtained from 19 BLAPs can be used to compare predicted and observed periods for the sample as a whole, not only on a star-to-star basis. From the theoretical relationship between $P$ and $g$ of Equation~(3), we would expect a slope $a$ of $-1.33$ in Equation~(4). Our observations, however, return a less steep relationship with $a=-1.14\pm0.05$. In Figure~\ref{fig:pg_models}, we compare our measurements with the periods predicted for the low-mass pre-WD evolutionary models of \citet{2019ApJ...878L..35K} shown in Figure~\ref{fig:param}. The orange lines, one for each mass from 0.28 to 0.35~\Msun, form a stripe defining the predicted periods of the fundamental mode. We also show the predicted periods for the first overtone with turquoise lines. Although the stellar mass affects the period, essentially by changing the intercept $b$ of the curve, for the range of masses involved in the low-mass pre-WD models, this effect is smaller than the typical error on $\log g$. The linear relation fitted to the theoretical periods, all masses included, for the fundamental mode is also indicated in Figure~\ref{fig:pg_models}. The slope obtained, $a=-1.309$, is very close to the value expected from the approximated analytical equation. Even though the predicted and observed slopes are different, our observations are nevertheless well reproduced by the models. We also included in Figure~\ref{fig:pg_models} the fundamental periods predicted for stellar models with higher masses ($M=0.5-0.8$~\Msun; \citealt{2023NatAs...7..223L}). In heavier objects, the fundamental periods are shifted to larger values for a given surface gravity and they are also consistent with our measurements. As mentioned in Section~\ref{sec:pg}, only the longest-period object, OGLE-BLAP-022, significantly deviates from the theoretical expectations. We note that this star also has a different light-curve shape in comparison to the remaining BLAPs (see Section~\ref{sec:photo} and Figure~\ref{fig:curves}). The higher $\log g$ than expected from the relation indicates that it cannot be an overtone pulsator. The detection and spectroscopic observations of other long-period ($P > 60$ minutes) BLAPs would help to verify the presence of a possible break in the linear relation. 

The comparison between the observations and the theoretical models suggests that the majority of BLAPs pulsate in the fundamental mode. We note here that the precision of the observed period--gravity relationship is strongly limited by our uncertainties on the $\log g$ measurements. As shown in previous works \citep{2017NatAs...1E.166P,2023NatAs...7..223L} and mentioned in Section~\ref{sec:atmo}, the surface gravity is changing during the pulsation cycle. The use of spectra averaged over one pulsation cycle, as opposed to time-resolved spectroscopy, as done by \citet{2019ApJ...878L..35K} for the high-gravity BLAPs and \citet{2024MNRAS.52710239B} for OGLE-BLAP-009, brings an extra uncertainty on the $\log g$ value derived from the spectral fits. To investigate this effect on our derived parameters, we fitted the individual low-resolution spectra of OGLE-BLAP-009 from \citet{2024MNRAS.52710239B} and obtained a log~$g$ value of 4.28 when averaging the measurements taken over a full pulsation period, only excluding the value from the spectrum taken at maximum luminosity (around phase 0.25 in Figures~1 and 2 of \citealt{2024MNRAS.52710239B}). We then coadded these spectra without applying any RV correction to mimic the effect of observing during a whole pulsation cycle. From the coadded spectrum we derived a surface gravity of log~$g$ = 4.36 $\pm$ 0.05. This shows that the systematic effect on the surface gravity due to the unresolved changes happening in the atmosphere of the star during the pulsation cycle should be lower than 0.1 dex.

\subsection{Evolutionary Status}

From their position in the \teff$-$log~$g$ diagram, and the $P$-$L$ relation, the majority of BLAPs could be low-mass pre-WDs. However, the peculiar pulsation properties of TMTS-BLAP-1 ($P=18.93$ minutes) point toward a different evolutionary origin. \citet{2023NatAs...7..223L} found that the large period change rate of TMTS-BLAP-1 is not compatible with the predictions from low-mass pre-WDs models and instead favors a different evolutionary channel, that of a more massive object ($\sim$0.7~\Msun) in a He-shell burning phase. The multimode pulsator OGLE-BLAP-030 could be a similarly massive object.

The period--gravity relations for the fundamental mode of the heavier ($M=0.5-0.8$~\Msun) models from \citet{2023NatAs...7..223L} are essentially parallel and slightly below that of the fundamental mode of the pre-extremely low-mass (pre-ELM) WD models (see Figure~\ref{fig:pg_models}). Given the uncertainties on the $\log g$ of our BLAPs, both types of models, low-mass pre-WDs or higher-mass He-shell burning objects, reproduce similarly well the observed periods for the BLAPs with $P\gtrsim 12$ minutes.

Although the low-mass pre-WD models reproduce well some characteristics of the BLAPs, they predict larger period change rates than what is observed \citep{2018MNRAS.477L..30R}. In addition, pre-WDs are contracting along their evolutionary tracks, which produces a negative $r$ value as the period decreases with time (see Equation~(1)). However, three-fourths of the BLAPs have a positive $r$ value which is difficult to reconcile with the low-mass pre-WD evolution. Both positive and negative $r$ were predicted by the models of \citet{2022A&A...668A.112X}. The authors computed helium-burning evolutionary models for a variety of masses (0.45--1.0~\Msun) that would reach parameters similar to those of the BLAPs (in terms of $\log g$ and \teff) during their He-shell burning, post-horizontal branch evolutionary phase. 

Another evolutionary channel worth discussing is that of \citet{2023ApJ...959...24Z}, who put forward that BLAPs can be explained by the merger of a He-core WD with a low-mass ($\sim$0.6~\Msun) main-sequence star. When the resulting object undergoes He-shell flashes, before settling into full helium-core burning, it has properties typical of BLAPs (in terms of \teff, log $g$, $L$ and helium abundances). The pulsational properties of BLAPs are also reproduced by these postmerger models, especially the rates of period change ($r$), which can be negative or positive following the cyclic expansion and contraction of He-shell burning stars. This formation scenario also naturally explains the He-enrichment observed in the low-gravity (i.e., long-period) BLAPs and predicts the stars with a surface gravity higher than log $g \sim$ 5.2 to be He-depleted. Their postmerger stars have masses in the range of 0.48--0.54~\Msun\ and the pulsation periods of the fundamental radial mode predicted by these models also agree with our observations.

Finally, we can compare some observed properties of BLAPs with the properties predicted by the evolutionary scenario in which these stars are possible surviving companions of single-degenerate Type Ia supernovae \citep{2020ApJ...903..100M}. In this scenario, the rates of period changes are of the order of $10^{-7}$ yr$^{-1}$ and they can be positive as well as negative. This is observed. The modeled companion stars have space velocities between 100 and 200 km~s$^{-1}$ relative to the center of mass of the binary systems. The measured RV values seem to reflect this range.

\subsection{Galactic Membership}

A Galactic bulge membership for the majority of the He-rich BLAPs discussed in Section~\ref{sec:kinem} is consistent with the large scatter in RVs that we measured in our spectroscopic sample. The values scatter from $-$200 to $+$200 km~s$^{-1}$, which is much larger than our typical uncertainties (10$-$20 km~s$^{-1}$). This is consistent with the RV dispersion of bulge members (about 140 km~s$^{-1}$) versus that of foreground Galactic disk stars \citep[$\sigma\sim40$ km~s$^{-1}$,][]{2018A&A...616A..83V}.

Although we could not measure directly the metallicity of the He-rich BLAPs from iron lines, we found numerous lines of singly or doubly ionized elements such as carbon, nitrogen, oxygen, magnesium, aluminum, and silicon. Nitrogen is the most enhanced element, with an abundance of about 10 times the solar value. The abundances of the other elements are 2--5 times higher than in the Sun. It is thus very likely that the overall metallicity of the He-rich BLAPs is supersolar. Such an enhanced metallicity was also recently determined for OGLE-BLAP-009 \citep{2024MNRAS.52710239B}. The authors used medium-resolution time-resolved spectroscopic observations of this bright star to derive elemental abundances for 10 metallic species, including Fe, and found them to be enhanced by factors between 2 and 5 as compared to the solar abundances. Interestingly, nitrogen is also the most enhanced species in OGLE-BLAP-009 with an abundance of about 25$\times$ solar. If an enhanced metallicity is key in their formation process, or for their pulsational instability, this would explain the absence of BLAPs in low-metallicity environments such as globular clusters, the Galactic halo, and the Magellanic Clouds \citep{2018pas6.conf..258P}.

\section{Summary} \label{sec:conc}

We reported the detection of 23 additional BLAPs to the first such pulsators found in the OGLE fields toward the inner Galactic bulge. We obtained moderate-resolution spectra of 11 BLAPs from the inner Galactic bulge fields and three BLAPs from the Galactic disk fields. Our analysis of the available data showed that the BLAPs form a slightly more complex class of objects than originally thought. Here, we summarize the observational facts on BLAPs.

BLAPs were discovered thanks to: (1) observations of a huge number of stars (400 million objects) toward the Milky Way's bulge, (2) the large amplitude of brightness variations ($>0.1$ mag), and (3) slow period changes (of an order of $10^{-7}$ to $10^{-6}$ yr$^{-1}$). Spectroscopically confirmed BLAPs cover pulsation periods between 3.3 and 75 minutes. All but two stars show only one periodicity in their power spectrum, and thus they are evident single-mode pulsators. Our analysis suggests that the BLAPs pulsate in the fundamental radial mode. One of the two unusual stars is the prototype, OGLE-BLAP-001, which exhibits a symmetric triplet likely due to rotational splitting with a frequency of $\nu_{\rm rot} \sim 1.903$ day$^{-1}$. The other star, OGLE-BLAP-030, is a triple-mode pulsator with significant amplitude and period changes. This star very likely pulsates in the fundamental mode, first overtone, and a nonradial mode. Typically, BLAPs have period change rates of few times $10^{-7}$ yr$^{-1}$. Only a quarter of the sample has a negative period change rate. In general, the changes are irregular on timescales of months to years, but in some objects, such as OGLE-BLAP-007, 019, and 022, secular trends seem to emerge.

A characteristic feature of a BLAP light curve is the sharp maximum. BLAPs with periods of 20--40 minutes have light-curve shapes similar to RRab-type stars, while those with periods below 20 minutes are more symmetric and rounded. In about half of the stars with periods longer than 40 minutes, an additional sinusoidal bump appears. The light curves of all objects but the mentioned OGLE-BLAP-030 are stable over years or even decades. No eclipses are seen in the 37 stars monitored by OGLE. This result practically rules out the possibility that an important fraction of BLAPs are in close binary systems.

The measured effective temperatures (\teff) of BLAPs are between 25,000 and 34,000 K, and their surface gravities ($\log g$) are between 4.2 and 5.7. This places the objects between the hot subdwarfs and upper main-sequence stars in the Hertzsprung-Russell diagram. We noticed a dichotomy in the helium-to-hydrogen content. Most of the analyzed stars group around \nhe\ $=-0.5$ dex, but the remaining stars group at \nhe\ $<$ $-2$ dex. There appears to be a transition in the atmospheric composition of the stars, such that the shortest-period BLAPs ($P \lesssim 15$ minutes), including the high-gravity BLAPs, have a He-poor atmosphere. These He-poor BLAPs are mostly found on the hottest end (\teff\ $>31,000$~K) of the temperature range, except for the newly discovered SMSS-BLAP-1 \citep{2024MNRAS.529.1414C}.

We showed that BLAPs follow a linear relationship between their pulsation period and surface gravity. For the first time in the case of Z-bump pulsators, we derived a $P$-$L$ relation. Such a relation is a powerful tool in estimating distances to the stars and verifying their membership status to stellar structures like clusters or nearby galaxies.

The majority of BLAPs observed in the OGLE Galactic bulge fields seem to reside in the bulge, while the remaining stars are in the Galactic disk. The closest object in our sample, OGLE-BLAP-009, is located at a distance of $2.2\pm0.2$ kpc, but the closest of all known BLAPs forms a binary system (HD 133729) with a main-sequence B-type star located at only about $463\pm7$ pc from the Sun. Detection of other Galactic BLAPs and more accurate distance determination should provide us the answer to the question of the true distribution of these stars in the Milky Way, that is, whether they concentrate around the Galactic center or rather along the Galactic disk. This should also give a hint as to the origin and evolutionary status of the stars.

There is a need for further spectroscopic observations and characterization of BLAPs to measure accurately their metallicity, to understand the transition in the helium abundance, and to verify if long-period BLAPs (with $P>40$ minutes) obey the same period--gravity relationship as short-period pulsators. More accurate distance measurements to OGLE-BLAP-009 and other objects of this type are expected from future Gaia data releases and will make it possible to improve the $P$-$L$ relation.

\begin{acknowledgments}
We thank all the OGLE observers for their contribution to the collection of the photometric data over the decades. We are grateful to P. N\'emeth for computing, and sharing with us, the atmospheric parameters of the combined spectrum of TMTS-BLAP-1; to C. Bradshaw for sharing the LRIS spectra of OGLE-BLAP-009; and to E. Bauer for providing us with the pre-ELM WD evolutionary tracks. We also thank X. Zhang and C. Byrne for sharing with us their evolutionary models to compare with our results, although they did not make it into our final figures. M.L. acknowledges funding from the Deutsche Forschungsgemeinschaft (grant LA 4383/4-1). P.P. has been supported by the Polish IDUB "Nowe Idee 3B" grant and "Microgrants" from the University of Warsaw, Poland. We used data from the European Space Agency (ESA) mission Gaia, processed by the Gaia Data Processing and Analysis Consortium (DPAC). Funding for the DPAC has been provided by national institutions, in particular, the institutions participating in the Gaia Multilateral Agreement. This research has made use of NASA’s Astrophysics Data System Bibliographic Services.
\end{acknowledgments}


\bibliography{paper}{}
\bibliographystyle{aasjournal}



\end{document}